\documentclass[10pt,a4paper,openright,tablecaptionabove]{scrartcl}
\usepackage{subfig}
\usepackage[english]{babel}
\usepackage{amsmath, amsthm, amssymb}
\usepackage{bbm}
\usepackage{url}
\usepackage{tikz-cd}
\usepackage{braket}
\usepackage{path}
\usepackage{nicefrac}
\usepackage{amssymb}
\usepackage{amsmath}
\usepackage{cite}
\usepackage{amsthm}
\usepackage{cancel}
\usepackage{indentfirst}
\usepackage{eucal}
\usepackage{appendix}
\usepackage[utf8]{inputenc}
\usepackage{geometry}
\usepackage{verbatim}
\usepackage{nameref}
\usepackage{alltt}
\usepackage[colorlinks=true, linkcolor=violet, citecolor=violet, urlcolor=violet]{hyperref}
\usepackage{pgfplots}
\usepackage{dsfont}
\usepackage{appendix}
\usepackage[font=small,labelfont=bf]{caption} 

\newtheorem{prop}{Proposition}

\theoremstyle{remark} 
\newtheorem{rem}{Remark} 

\newcommand{\id}{\hat{\mathds{1}}}

\newcommand{\hbx}{\hat{\bx}}
\newcommand{\bx}{\boldsymbol{{x}}}
\newcommand{\hx}{\hat{{x}}}

\newcommand{\hbp}{\hat{\boldsymbol{p}}}
\newcommand{\bp}{\boldsymbol{{p}}}
\newcommand{\hp}{\hat{{p}}}

\newcommand{\hbr}{\hat{\boldsymbol{R}}}

\newcommand{\hr}{\hat{R}}

\newcommand{\hbpi}{\hat{\boldsymbol{\pi}}}

\newcommand{\hpi}{\hat{{\pi}}}

\newcommand{\hl}{\hat{\Lambda}}

\newcommand{\D}{\text{D}}

\newcommand\be{\begin{equation}}
\newcommand\ee{\end{equation}}

 \newcommand{\bR}{ \mathbb{R}} 
\newcommand{\bE}{ \mathbb{E}} 
\newcommand{\bS}{ \mathbb{S}} 
\newcommand{\bH}{ \mathbb{H}} 
 \newcommand{\bD}{ \mathbb{D}_\text{III}} 

\newcommand{\al}{\rho} 
\newcommand{\ii}{\alpha} 

\newcommand{\kk}{m} 
\newcommand{\ff}{f} 
\newcommand{\dd}{\text{d}} 
\newcommand{\RR}{\mathcal{R}} 

 \newcommand{\gga}{\Psi} 
  \newcommand{\ggb}{\Phi} 
 \newcommand{\mani}{\mathcal{M}}

  \newcommand{\amb}{s}
  \newcommand{\ambb}{\boldsymbol{s}} 
   \newcommand{\metric}{s}
 \newcommand{\parp}{\xi}

  \newcommand{\TT}{T}
 
  \newcommand{\yy}{y}
  \newcommand{\byy}{\boldsymbol{y}}

\usepackage{color}

\areaset[current]{500pt}{680pt}
\addtokomafont{disposition}{\normalfont\bfseries}

\numberwithin{equation}{section}
\begin{document}

\ 
\bigskip

\thispagestyle{empty}
\begin{center}
\Large{\textbf{An infinite family of Dunkl type superintegrable  curved Hamiltonians through coalgebra symmetry: Oscillator and Kepler-Coulomb models}}
\end{center}
\vskip 0.5cm
\begin{center}
	\textsc{Francisco J. Herranz$^{1,*}$ and Danilo Latini$^{2,3,\star}$}
\end{center}
\begin{center}
	$^1$ Departamento de F\'isica, Universidad de Burgos, E-09001 Burgos, Spain 
\end{center}
\begin{center}	
	$^2$ Universit\`a degli Studi di Milano, Dipartimento di Matematica ``Federigo Enriques", \\
	Via Cesare Saldini 50, 20133, Milano, Italy
\end{center}
\begin{center}
	$^3$ INFN Sezione di Milano, Via Giovanni Celoria 16, 20133 Milano, Italy 
\end{center}
\begin{center}
	\footnotesize{
		$^*$\textsf{fjherranz@ubu.es}  \hskip 0.25cm $^\star$\textsf{danilo.latini@unimi.it} }
\end{center}
\vskip 0.75cm
\begin{center}
{\bf Abstract}
\end{center}
 \begin{abstract}
 \noindent  This work aims to bridge the gap between Dunkl superintegrable systems and the coalgebra symmetry approach to superintegrability, and subsequently  to recover known models and construct new ones. In particular, an infinite family of $N$-dimensional quasi-maximally superintegrable quantum systems with reflections, sharing the same set of  $2N-3$ quantum integrals, is introduced. The result is achieved by introducing a novel differential-difference realization of  $\mathfrak{sl}(2, \mathbb{R})$ and then applying the coalgebra formalism.  Several well-known maximally superintegrable models with reflections appear as particular cases of this general family, among them, the celebrated Dunkl oscillator and the Dunkl-Kepler-Coulomb system. Furthermore, restricting to the case of  ``hidden" quantum quadratic symmetries,  maximally superintegrable curved oscillator and Kepler-Coulomb Hamiltonians of Dunkl type, sharing the same underlying $\mathfrak{sl}(2, \mathbb{R})$ coalgebra symmetry, are presented. Namely, the Dunkl oscillator and the Dunkl-Kepler-Coulomb system on the $N$-sphere and hyperbolic space together with two models which can be interpreted as a one-parameter superintegrable deformation of the Dunkl oscillator and the Dunkl-Kepler-Coulomb system  on  non-constant curvature spaces. In addition, maximally superintegrable generalizations of these models, involving non-central potentials, are also derived  on flat and curved spaces. 
  For all  specific systems, at least an additional quantum integral is explicitly provided, which is related to the Dunkl version of a (curved)    Demkov-Fradkin tensor or a Laplace-Runge-Lenz  vector.
  
 \end{abstract}
\vskip 0.35cm
\hrule

\medskip

\medskip

\bigskip
\hrule

\newpage

\tableofcontents


\section{Introduction}
\label{intro}
\noindent Maximally superintegrable systems represent a distinct subset of finite-dimensional integrable Hamiltonian systems, distinguished by their unusually high number of symmetries. A classical Hamiltonian system 
$H=H(\bx,\bp)$ with $N$ degrees of freedom is Liouville integrable if it is endowed with $N$ functionally independent, analytic and single-valued constants of motion $I_i$, with $i=1, \dots, N$ and $I_1:=H$, which are in involution, i.e.~they Poisson commute $\{I_i,I_j\}=0$. In this case, the Liouville theorem \cite{Liouville1853, whittaker} guarantees that the system of  $2N$ canonical differential equations can be integrated by quadratures. Moreover, (bounded) invariant manifolds defined by the equations $I_i = \alpha_i = \text{constant}$ $(i = 1, \dots, N)$ turn out to be tori, and motions on these tori are conditionally periodic \cite{Arnol'd1997}. If, in addition to the $N$ first integrals, there exist other $k$ functionally independent constants of motion, with $1\leq k \leq N-1$, the system is called superintegrable (or, degenerately integrable \cite{Reshetikhin}). In this case, the dimension of the invariant tori is less than $N$ (see \cite{Fasso} for a comprehensive characterization of the geometry of superintegrable systems). When $k=1$, they are minimally superintegrable systems, whereas when $k=N-1$   they are said to be maximally superintegrable (MS). Remarkably, in the MS case, all finite trajectories are closed, and the motion is periodic \cite{Nek1972}.
As a matter of fact, MS systems are multi-integrable, meaning that there can exist multiple subsets of 
$N$ involutive constants of motion among the total $2N-1$. Several celebrated models belong to the class of MS systems, including the isotropic harmonic oscillator, the anisotropic harmonic oscillator (with commensurate frequencies) \cite{Jauch1940,Stefan2002,Tempesta2008,Kuru2016}, the Kepler-Coulomb system, the Calogero-Moser system \cite{Woj1983}, and the Toda lattice \cite{ADS2006}, just to name a few.
Regarding the quantum mechanical analogues of superintegrable systems, MS property, defined as the existence of $2N-1$ algebraically independent quantum integrals (realized as finite-order partial differential operators), is closely related to exact solvability \cite{TTW2001, RW2002, Fordy2007}. Furthermore, in such a highly constrained case, the models show maximal degeneracy of the energy levels.  For a historical overview and a state-of-the-art perspective on superintegrable systems, we refer the reader to the papers \cite{MPW, BBM}. 

If only one constant is missing to achieve a classical MS system, meaning that there is a total number of  $2N-2$ functionally independent constants, the system is typically referred to as quasi-maximally superintegrable (QMS). A well-known example of QMS system in $N$ dimensions is the general central force problem, where the Hamiltonian is given by $H=H(\bx, \bp)=\bp^2/2+V(r)$, with $r:=|\bx|=\sqrt{\bx^2}$. In this case, rotational invariance implies the existence of a subset composed by $2N-3$ functionally independent constants of motion, derived as suitable quadratic combinations of the $N(N-1)/2$ (angular momentum) $\mathfrak{so}(N)$ generators  $L_{ij}=x_i p_j-x_j p_i$.
Among central force potentials, there are two specific cases that are MS, namely
the harmonic oscillator potential $V_{\textsf{HO}}(r)= \omega^2 r^2/2$ and the Kepler-Coulomb (KC) potential $V_{\textsf{KC}}(r)=-k/r$.
In three dimensions, this result is a consequence of Bertrand's Theorem \cite{Bertrand1873}, which states that, among all central force potentials in three-dimensional Euclidean space, only two cases lead to all bounded orbits being closed: the oscillator and the KC potentials mentioned above.
For these two systems, MS is guaranteed by the existence of two additional constants of motion (quadratic in the momenta), which are the (symmetric) Demkov-Fradkin  tensor \cite{D1959, F1965}, whose $N(N+1)/2$ components $F_{ij}=p_ip_j+\omega^2 x_i x_j$ are conserved for $V=V_{\textsf{HO}}(r)$ and the Laplace-Runge-Lenz  (LRL)
$N$-vector \cite{G1975, G1976, GPS2002}, whose $N$ components $A_i=\sum_{j=1}^N L_{ij}p_j-k {x_i}/{r}$ are conserved for $V=V_{\textsf{KC}}(r)$. Interestingly enough, when non-central terms of the type $\beta_i/x_i^2$ are added to the central force Hamiltonian above, QMS property still survives thanks to the existence of $2N-3$ functionally independent constants of motion derived from the $N(N-1)/2$  integrals of the type: $K_{ij}=L_{ij}^2+\beta_i x_j^2/x_i^2+\beta_j x_i^2/x_j^2$, with $K_{ij}=K_{ji}$. As a matter of fact, it has been proved that the family of classical $N$-dimensional ($N$D) Hamiltonians of the type $H=F\left(\bp^2+\sum_{i=1}^N \beta_i/x_i^2, \bx^2, \bx \cdot \bp\right)$, for  suitably well-behaved choices of the function $F$, turns out to be QMS \cite{BHMR2004,BH2007}. This is because, in addition to the Hamiltonian, there exist  $2N-3$ quadratic integrals of motion. These are referred to as ``universal" integrals because they are shared by the entire family of Hamiltonians. This result also holds in the quantum setting, where the corresponding family of Hamiltonians and the associated universal constants of motion have their quantum counterparts \cite{Latini2019}.
A detailed explanation of this result can be found within the framework of the coalgebra symmetry approach to superintegrability \cite{BalCorRag1996, BalRag1998, BalBlaHerMusRag2009}, where the $2N-3$ functionally independent constants of motion naturally arise as the image, under a given symplectic realization, of the so-called left and right Casimirs of the $\mathfrak{sl}(2,\mathbb{R})$ coalgebra. In the case of spherically symmetric systems, these appear as the quadratic Casimirs associated with certain rotation subalgebras of $\mathfrak{so}(N)$, while in the presence of non-central terms, they manifest as suitable linear combinations of the $K_{ij}$, which are instead related to the generalized Racah algebra $R(N)$ \cite{Latini2019, DIVV2021, LMZ2021, LMZ2021b}.
The general idea behind this approach to (super)integrability is to reinterpret 1D dynamical Hamiltonian systems as the images, under a given realization, of some (smooth) function of the coalgebra generators. Using the coproduct map, the method extends the $1$D system to higher dimensions in order to obtain a multi-dimensional version of the original Hamiltonian. The key point is that the higher-dimensional system will possess, by construction, constants of motion arising from the aforementioned left and right Casimirs of the coalgebra. At a fixed realization, these generate two pyramidal sets composed of $N-1$ constants in involution, with one constant  being in common. The union of these two sets produces a unique set consisting of $2N-3$ functionally independent constants of motion.

This construction has also been used (in relation to different types of problems) in the quantum case, where Poisson brackets are replaced by commutators and symplectic realizations are replaced by realizations of the Lie algebra in terms of differential operators \cite{Latini2019, LMZ2021, LMZ2021b, BEHRR2011, Riglioni2013, RGW2014, PostRiglioni2015}. Moreover, the method has found applications not only in the continuous setting but also in the discrete one, both in classical \cite{GLT2023, GL2023, GD2024} and quantum mechanics \cite{LR2016}.  
 In addition, it is worth noting that since its introduction  \cite{BalCorRag1996, BalRag1998,BalBlaHerMusRag2009}, the same coalgebra formalism has also been used with quantum groups or  $q$-deformations, which have been applied to obtaining quantum group deformations of several relevant (dynamical) systems  (see, e.g., \cite{BalBlaHerMusRag2009,Chains1999,Musso2000,ballesteros2003cluster,BHR2005,BGSMR2025} and references therein for specific $q$-deformed Hamiltonians in different contexts). The widespread application of this technique is due to its purely algebraic nature. The general results are, in fact, obtained algebraically, and the connection with different frameworks is established through suitable realizations, which can be symplectic, differential, difference, etc.

Taking that into account, among the class of superintegrable quantum systems, there exist several works  in the literature devoted to the so-called Dunkl type superintegrable systems \cite{GIVZ2013I,GIVZ2013II, GVZ2013, GVZ2014,GLV2015, GSAE2019, GS2020, Ghaz2021, FH2022, NajaPana, DNPCH,BNPHD2024, Quesne2024, GM2025}. These involve the presence of differential-difference operators introduced by C. Dunkl in \cite{Dunkl1989}. These operators are associated to finite reflection groups \cite{hump1990} and can be thought as a generalization of the standard derivatives \cite{Ros2003, Dunkl2012}. Roughly speaking, these superintegrable models are obtained by replacing the momentum operator with its Dunkl counterpart. This causes a deformation of the original Hamiltonians to accomodate reflection operators. Interestingly enough, such a replacement in many cases keep the models superintegrable, by providing an extension of the quantum integrals to the Dunkl realm.

The aim of this paper is twofold. Firstly, the main  objective is to link the above-mentioned coalgebra symmetry method to superintegrability within the framework of Dunkl superintegrable systems. This is performed by providing a novel differential-difference realization of the Lie algebra $\mathfrak{sl}(2,\mathbb{R})$ as extended Dunkl operators associated to the $\mathbb{Z}_2^N$ reflection group, which is then  used to characterize the coalgebraic Hamiltonians of Dunkl type.  Within the coalgebra symmetry procedure, we are able to introduce an infinite family of $N$D Hamiltonians of Dunkl type which is  QMS in any dimension $N$. Secondly, we extensively apply the previous result to construct  and discuss specific MS systems  by deriving additional ``hidden" quantum symmetries to those of the generic  QMS family, namely the corresponding Dunkl-Demkov-Fradkin tensor or  Dunkl-LRL vector (or some of their components).
In particular,  we recover well-known Dunkl superintegrable models, such as for example the celebrated Dunkl oscillator \cite{GIVZ2013I,GIVZ2013II, GVZ2014, Ghaz2021} or the Dunkl-KC system \cite{GLV2015,GSAE2019, GS2020, Ghaz2021,FH2022,Quesne2024,GM2025}, which are here understood as systems possessing a common $\mathfrak{sl}(2,\mathbb{R})$ coalgebra symmetry.   Moreover, other  MS subcases   are identified and studied, specifically, 
the   Dunkl oscillator on the $N$D sphere $\bS^N$ and hyperbolic space $\bH^N$, already known for $N=2$~\cite{NajaPana} and $N=3$~\cite{DNPCH},    the  new Dunkl-KC system on such spaces of constant curvature,  as well as two models which can be interpreted as a one-parameter deformation of the Dunkl oscillator and Dunkl-KC system on spaces of non-constant curvature. The former has been introduced for the first time in the recent paper  \cite{BNPHD2024} on the so-called Darboux III space  \cite{BEHRR2011,Latini2016} and, again, it is here understood as a system with $\mathfrak{sl}(2,\mathbb{R})$ coalgebra symmetry. The latter can be instead considered as a novel MS extension of the Dunkl-KC system on the Taub-NUT space \cite{BEHRR2014,LatiniNUT}. Additionally, as a consequence of the differential-difference realization of the $\mathfrak{sl}(2,\mathbb{R})$ coalgebra,   MS  generalizations of all  these models are also obtained, which are understood as the superposition of the   oscillator or KC potential with  non-central terms, both on flat and curved spaces.  To  our knowledge, they have so far only been considered   for  the   Dunkl oscillator on the Euclidean plane $\bE^2$ in~\cite{GVZ2013}.
 
 \medskip

  The paper is organized as follows:
\begin{itemize}
\item  In Section \ref{sec2} we present the main result of the paper, which is established in Proposition~\ref{prop1}. In particular, after reviewing some notions about Dunkl operators associated to the $\mathbb{Z}_2^N$ reflection group, we proceed by introducing the infinite family of QMS systems of Dunkl type. We show, by making use of the coalgebra symmetry method, that the $2N-3$ Dunkl quantum integrals arise as the left and right partial Casimirs of the coalgebra $\mathfrak{sl}(2,\mathbb{R})$ through a newly introduced differential-difference realization involving reflection operators.

\item The generic results thus obtained  are then applied in Section \ref{sec3} to construct the most general family of $N$D QMS quantum Hamiltonians of Dunkl type with central potentials on spherically symmetric spaces which are,  in general, of non-constant curvature.  Therefore, the corresponding systems on the  sphere $\bS^N$, Euclidean $\bE^N$ and hyperbolic  $\bH^N$ spaces arise as the particular cases with constant curvature.
 
  \item Section \ref{sec4} is devoted to identify and  study  six specific MS Hamiltonians, namely Dunkl oscillator-type systems.  We begin the discussion by showing that many MS  oscillators  of Dunkl type (some appearing in the literature and some completely new) are in fact coalgebraic, meaning that they share the same underlying $\mathfrak{sl}(2,\mathbb{R})$ coalgebra symmetry. They include the well-known proper Dunkl oscillator \cite{GIVZ2013I,GIVZ2013II, GVZ2014, Ghaz2021}, the generalized  or singular Dunkl oscillator  \cite{GVZ2013}   (here intrepreted as the Dunkl-Smorodinsky-Winternitz), the Dunkl (Higgs) oscillator on $\bS^N$ and $\bH^N$    \cite{NajaPana,DNPCH}  together with their novel generalization involving non-central potentials. In addition, we also present the recently introduced Dunkl-Darboux III system  \cite{BNPHD2024}, which extends  the Darboux III oscillator (i.e., on a space of non-constant curvature) \cite{BEHRR2011, Latini2016} to the Dunkl framework and, also  in this case, we construct its new generalization through $N$ non-central terms, thus extending to the Dunkl setting the results obtained in~\cite{BEHR08a}.  For each model, the corresponding additional Dunkl quantum integrals are explicitly shown, which are quadratic in the momenta. These can result in a tensor of Dunkl-Demkov-Fradkin type or,  for most systems,   an extension of its $N$ diagonal components.
   
\item In a ``parallel" way to the previous oscillator systems, we deduce in Section \ref{sec5} six MS Dunkl-KC-type Hamiltonians. We remark that, with the exception of the   well-known Dunkl-KC system  \cite{GLV2015,GSAE2019, GS2020, Ghaz2021,FH2022,Quesne2024,GM2025}, all the models presented are new.  In particular, they cover the   Dunkl version of the quasi-generalized KC system  \cite{Evans90b,KWMP1999, KWMP2002, RW2002, LMZ2018}, the Dunkl-KC system on $\bS^N$ and    $\bH^N$ together with their generalization involving $N-1$ non-central terms \cite{KC2009}. 
As a last application, we consider a Dunkl version of the Taub-NUT system \cite{ BEHRR2014,LatiniNUT}, together with its quasi-generalized counterpart.  Furthermore, also in this case, for each model, the corresponding quadratic Dunkl quantum integrals are given, which either provide an  $N$-vector  of Dunkl-LRL type or a single component. 

\item Finally, some remarks and  open perspectives close the paper. 

\end{itemize}

\noindent It is worth remarking that the key tool for deriving quantum integrals that reveal the MS property of Dunkl oscillator and KC systems on $\bS^N$ and $\bH^N$ is the construction of appropriate curved spherical/hyperbolic Dunkl momenta in these spaces. The corresponding algebraic structure is discussed in Appendix~\ref{AA}.

 \noindent  We also note that throughout the paper the terminology ``generalized" system refers  to the addition of $N$ arbitrary non-central potentials to the main oscillator/KC potential, while ``quasi-generalized" system means that only  $N-1$ non-central terms are added. Since we restrict ourselves to deriving   quadratic integrals which ensure the MS property, it follows that for oscillator type systems we can always consider $N$ non-central potentials, whereas for KC type models, we can only introduce $N-1$ non-central terms. The reason behind this fact is that  generalized KC systems are endowed with quartic integrals, even in the simplest (flat) Euclidean space \cite{KC2009,VE2008,TD2011}.


\section{An infinite family of quasi-maximally superintegrable systems of Dunkl type in $\boldsymbol{N}$ dimensions}
\label{sec2}

\noindent
 In this section, after presenting the basic notions about Dunkl operators associated with the $\mathbb{Z}_2^N$ reflection group, which are needed to state the main result of the work, we construct a new infinite family of  $N$D QMS systems of Dunkl type by making use of the coalgebra symmetry technique. The  key  link between Dunkl superintegrable systems and the coalgebra symmetry approach to superintegrability is provided by a novel differential-difference realization of $\mathfrak{sl}(2,\mathbb{R})$,  which allows us to extend this algebraic approach even in the presence of Dunkl operators. 


\subsection{Dunkl operators associated with the $\mathbb{Z}_2^N$ reflection group}
\label{subsec2.1}

\noindent 
Let us introduce all the necessary operators used throughout the paper. 
\begin{itemize}

\item \textbf{Position operators}: 
\begin{equation}
\hbx = (\hx_1 ,\dots, \hx_N ) \, ,  \quad \hx_i \Psi(\bx) = x_i \Psi(\bx)\, .
\label{positions}
\end{equation}
These operators satisfy the commutation relations: 
\begin{equation}
[\hat{x}_i, \hat{x}_j]=0 \, , \quad  i,j=1, \dots, N\, .
\label{poscomm}
\end{equation}
We will also need the squared position operator $\boldsymbol{\hx}^2=\sum_{i=1}^N \hx_i^2$, such as $\boldsymbol{\hx}^2\Psi(\bx)=\bx^2 \Psi(\bx)$. Then,  we denote $|\boldsymbol{\hx}|:= \sqrt{\hbx^2}$  and $|\boldsymbol{\hx}|\Psi(\bx)=|\bx |\Psi(\bx)$.

\item \textbf{Momentum operators}: 
\begin{equation}
\hskip 0.6cm\hbp = (\hp_1 ,\dots, \hp_N ) \, ,  \quad \hp_i \Psi(\bx) = - \imath \hbar \partial_i \Psi(\boldsymbol{\bx})\, ,
\label{momenta}
\end{equation}
where we have introduced the shorthand notation $\partial_i := \partial/\partial x_i$.
These operators satisfy the commutation relations:
\begin{equation}
 [\hat{p}_i, \hat{p}_j]=0\, , \quad i,j=1, \dots, N\, .
 \label{momcomm}
 \end{equation}
 Moreover, the commutation between position and momentum operators gives:
 \begin{equation}
 	[\hat{x}_i, \hat{p}_j]=\imath \hbar \delta_{ij} \hat{\mathds{1}}\, , \quad i,j=1, \dots, N\, ,
 	\label{momposcomm}
 \end{equation}
where $\hat{\mathds{1}}$ indicates the identity operator. The relations \eqref{poscomm}, \eqref{momcomm} and \eqref{momposcomm} define the ($2N+1$)D Heisenberg algebra $\mathfrak{h}_N$. At this level, the underlying Hilbert space is $L^2\big(\mathbb{R}^N, \text{d}\bx \big)$, with inner product:
\begin{equation}
\braket{\gga|\ggb}:=\int_{\mathbb{R}^N}\overline{\gga(\bx)}\ggb(\bx)\text{d}\boldsymbol{x}\, .
\label{eq:innprodstand}
\end{equation}

\item \textbf{Reflection operators}: 
\begin{equation}
\hskip 0.4cm \hbr = \big(\hr_1 ,\dots, \hr_N \big) \, ,  \quad \hr_i\Psi(\bx) =\Psi(\sigma_i (\bx)) \, ,
\label{reflections}
\end{equation}
with
\begin{equation}
\sigma_i(\bx) := \bx-2\,\braket{\boldsymbol{e}_i, \bx} \boldsymbol{e}_i \, .
\label{explicitreflection}
\end{equation}
Here $\{\boldsymbol{e}_i\}_{i=1}^N$ is the standard basis of the Euclidean vector space $\mathbb{R}^N$ and $\braket{\cdot, \cdot}$ is the standard Euclidean scalar product in $\mathbb{R}^N$. This represents a reflection of $\bx$  with respect to the hyperplane $\boldsymbol{e}_i^{\perp}$, orthogonal to $\boldsymbol{e}_i$. The explicit action on functions is given by
\begin{equation}
\hr_i\Psi(x_1, \dots, x_i, \dots, x_N) =\Psi(x_1, \dots, -x_i, \dots, x_N) \, , \quad i=1, \dots, N \, .
\label{action}
\end{equation}
These are reflection operators associated to the Abelian reflection group $\mathbb{Z}_2^N$  \cite{Dunkl1989, Ros2003}. These operators commute each other, i.e.:
\begin{equation}
  \big[\hr_i, \hr_j \big]=0 \, , \quad i,j=1, \dots, N \, .
\label{reflcomm}
\end{equation}
In addition, they satisfy the following commutation relations with the position and momentum operators:
\begin{equation}
\big[\hx_i, \hr_j \big]=2\delta_{ij}\hx_i \hr_j \, , \quad \big[\hp_i, \hr_j \big]=2\delta_{ij}\hp_i \hr_j \, , \quad i,j=1, \dots, N \, .
\label{reflposmomcomm}
\end{equation}

\item \textbf{Dunkl momentum operators}: 
\begin{equation}
\hbpi= (\hpi_1 ,\dots, \hpi_N ) = \left(\hp_1-\imath \hbar \frac{\mu_1}{\hx_1}\big(\id-\hr_1 \big), \dots, \hp_N-\imath \hbar\frac{\mu_N}{\hx_N}\big(\id-\hr_N\big)\right) \, , \quad \hpi_i \Psi(\bx) = -\imath \hbar \text{D}_i \Psi(\bx) \, ,
\label{dunklmom}
\end{equation}
where $\D_i$ is the Dunkl derivative along the direction of  $\boldsymbol{e}_i$, i.e.~the $i$-th component of the $N$D Dunkl gradient $N$-vector $\mathcal{\nabla}_\D=(\D_1, \dots, \D_N)$. Specifically, the action is the following:
\begin{equation}
\hpi_i \Psi(\bx) = -\imath \hbar \text{D}_i \Psi(\bx) = - \imath \hbar \left(\partial_i \Psi(\bx) + \frac{\mu_i}{x_i} \bigl(\Psi(\bx)-\Psi(\sigma_i(\bx) )\bigl) \right) \, ,
\label{eq:explmom}
\end{equation}
where $\{\mu_i\}_{i=1}^N$ are assumed to be real parameters (usually they are taken such as $\mu_i >-1/2$). The introduction of Dunkl derivatives requires a change of  the inner product \eqref{eq:innprodstand}, which has to be defined as (see, e.g., \cite{Quesne2024}):
\begin{equation}
\braket{\gga | \ggb}_{\boldsymbol{\mu}}:=	\int_{\mathbb{R}^N} \overline{\gga(\bx)}\ggb(\bx)\prod_{i=1}^{N}|x_i|^{2\mu_i}\text{d}\bx \, ,
\label{eq:inner}
\end{equation}
the corresponding Hilbert space being $L^2\!\left(\mathbb{R}^N, \prod_{i=1}^{N}|x_i|^{2\mu_i}\text{d}\bx\right)$, with 
\begin{equation}
\prod_{i=1}^{N}|x_i|^{2\mu_i}\text{d}\bx =|x_1|^{2\mu_1}\dots|x_N|^{2\mu_N}\text{d}x_1 \dots \text{d}x_N\, .
\end{equation}
As a consequence of the fact that Dunkl derivatives commute, these operators satisfy:
\begin{equation}
[\hpi_i,\hpi_j]=0 \, ,  \quad i,j=1, \dots, N \, .
\end{equation}
Furthermore, their commutation relations with position and reflection operators read:
\begin{equation}
[\hx_i, \hpi_j] = \imath \hbar \delta_{ij} \big(\id+2 \mu_j \hr_j\big) \, , \quad  \big[\hpi_i,\hr_j \big]= 2 \delta_{ij} \hpi_j \hr_j \, ,  \quad i,j=1, \dots, N \, .
\end{equation}
We also consider the operator:
\begin{equation}
\boldsymbol{\hpi}^2=\sum_{i=1}^N \hpi_i^2=-\hbar^2 \sum_{i=1}^N D_i^2=-\hbar^2\sum_{i=1}^N \left( \partial_i^2 +2\frac{\mu_i}{x_i}\partial_i -\frac{\mu_i}{x_i^2} \bigl(\id-\hr_i\bigl)\right)=-\hbar^2 \Delta_\D
\label{dunkllap}
\end{equation}
where we have introduced $\Delta_\D = \nabla_\D^2$, namely the Dunkl Laplacian in dimension $N$. The explicit action turns out to be:
\begin{equation}
\boldsymbol{\hpi}^2\Psi(\bx)=-\hbar^2\Delta_\D \Psi(\bx)=-\hbar^2\sum_{i=1}^N \left(\partial_i^2 \Psi(\bx)+2\frac{\mu_i}{x_i}\partial_i \Psi(\bx)-\frac{\mu_i}{x_i^2} \bigl(\Psi(\bx)-\Psi(\sigma_i(\bx))\bigl)\right) \, .
\label{eq: lapdun}
\end{equation}
We will also make use of the following operator:
\begin{equation}
\boldsymbol{\hx} \cdot \boldsymbol{\hpi}=\sum_{i=1}^N \hx_i \hpi_i=-\imath \hbar \sum_{i=1}^N x_i D_i=-\imath \hbar (\bx \cdot \nabla_\D) \, , 
\end{equation}
whose explicit action is given by
\begin{equation}
(\boldsymbol{\hx} \cdot \boldsymbol{\hpi} )\Psi(\bx)=-\imath \hbar (\bx \cdot \nabla_\D)\Psi(\bx)=-\imath \hbar \sum_{i=1}^N x_i D_i \Psi(\bx)=-\imath \hbar \sum_{i=1}^N \bigl(x_i \partial_i \Psi(\bx)+ \mu_i \bigl(\Psi(\bx)-\Psi(\sigma_i(\bx))\bigl)\bigl) \, .
\label{dpix}
\end{equation}

\item \textbf{Dunkl angular momentum operators}: 
\begin{equation}
\hl_{ij} = \hx_i \hpi_j - \hx_j \hpi_i \, , \quad \hl_{ij}  \Psi(\bx) = -\imath \hbar \bigl(x_i\D_j \Psi(\bx)-x_j \D_i \Psi(\bx)\bigl)  \, ,
\label{eq:angmomdunkl}
\end{equation}
which, to make explicit their dependence on the standard angular momentum operators, can be expressed as
\begin{equation}
\hl_{ij}  =\hat{L}_{ij}  +\imath \hbar \left( \mu_i \frac{\hx_j}{\hx_i}\big( \id-  \hr_i\big)-\mu_j \frac{\hx_i}{\hx_j}\big(\id-\hr_j \big) \right) 		\, ,
\label{momentaangdep}
\end{equation}
\noindent where 
\be
\hat{L}_{ij} =\hat{x}_i \hat{p}_j-\hat{x}_j \hat{p}_i\, .
\label{standardL}
\ee
 The explicit action can be therefore rewritten as:
\begin{equation}
	\hl_{ij}   \Psi(\bx)=\imath \hbar \left( x_j\partial_i \Psi(\bx)-x_i\partial_j \Psi(\bx)  +\mu_i \frac{x_j}{x_i}\bigl(\Psi(\bx)-\Psi(\sigma_i(\bx))\bigl)-\mu_j \frac{x_i}{x_j}\bigl(\Psi(\bx)- \Psi(\sigma_j(\bx))\bigl)\right) 	\, .
\label{expang}	
\end{equation}
We observe that these Dunkl angular momentum operators close  an extension by reflections of the Lie algebra $\mathfrak{so}(N)$, with commutation relations given by (for $i,j,k,l=1, \dots, N$):
\begin{equation}
\begin{split}
\big[\hat{\Lambda}_{ij}, \hat{\Lambda}_{kl} \big]&=\imath \hbar \left(\delta_{ik}\big(\id+2\mu_k \hr_k \big) \hat{\Lambda}_{jl}+\delta_{jl}\big(\id+2\mu_l \hr_l \big) \hat{\Lambda}_{ik}-\delta_{il}\big(\id+2\mu_i \hr_i \big) \hat{\Lambda}_{jk}-\delta_{jk}\big(\id+2\mu_j \hr_j \big) \hat{\Lambda}_{il}\right) \\
  \big[\hat{\Lambda}_{ij}, \hr_{k} \big]  & =2 \delta_{ik}\hat{\Lambda}_{ij}\hr_k+2 \delta_{jk}\hat{\Lambda}_{ik}\hr_j\ ,  
\end{split}
\label{eq:so(n)R}
\end{equation}
together with \eqref{reflcomm}. 
\end{itemize}

\begin{rem}
\label{rem1}
In the recent paper \cite{Ghaz2021}, the algebraic structure formed by the commutation relations  (\ref{reflcomm})  and (\ref{eq:so(n)R}) has been denoted  $\mathfrak{so}\big(N, \mu_1 \hr_1, \dots, \mu_N  \hr_N\big)$, thus referring to the quadratic algebra spanned by the $N(N-1)/2$ angular momentum generators $\hat{\Lambda}_{ij}$ ($1\le i<j\le N$) and   $N$ reflections $\hr_i$. Clearly, when reflections are discarded the quadratic algebra yields the usual  Lie algebra $\mathfrak{so}(N)$.  Note  that the non-zero  commutators in the second relation of (\ref{eq:so(n)R}) can  also be written in terms of anticommutators $\{\cdot, \cdot\}$, as in   \cite{Ghaz2021}, namely:
\be
\big\{ \hat{\Lambda}_{ij}, \hr_i  \big\}=\big\{ \hat{\Lambda}_{ij},\hr_j  \big\}=0\, , \quad \big[ \hat{\Lambda}_{ij} ,\hr_k \big]=0\, ,\quad i\ne j\ne k\, .
\ee
\end{rem}

\medskip

\noindent At this point, we have all the necessary ingredients to present the main result of the paper.


 \subsection{An infinite family of  QMS systems of Dunkl type through $\mathfrak{sl}(2,\mathbb{R})$  coalgebra symmetry}

\noindent Let us consider the  family of $N$D quantum Hamiltonians with reflections defined by
	\begin{equation}
	\hat{H} := F\left(\hbpi^2+\sum_{i=1}^N \frac{\beta_i +\gamma_i \hr_i}{\hx_i^2} \, , \, \hbx^2 \, , \, \hbx \cdot \hbpi-\imath \hbar \left( \frac N2 + \sum_{i=1}^N \mu_i \hr_i\right)    \right)\, ,
	\label{ham}
	\end{equation}
\noindent where  $F$ is an arbitrary function, $\{\beta_i, \mu_i, \gamma_i\}_{i=1}^N$ are real parameters, the operators $\hbx$, $\hbpi$, $\hbr$ satisfy the    commutators:
\begin{equation}
\begin{split}
&	[\hx_i, \hx_j]=[\hpi_i, \hpi_j]=\big[\hr_i,\hr_j \big]=0 \, , \quad \big[\hpi_i,\hr_j\big]= 2 \delta_{ij} \hpi_j \hr_j  \\[2pt]
&	[\hx_i, \hpi_j] = \imath \hbar \delta_{ij} \big(\id+2 \mu_j \hr_j \big) \, , \hskip 1.100cm \big[\hx_i,\hr_j\big]= 2 \delta_{ij} \hx_j \hr_j \, ,
\end{split}
\label{eq:2}
\end{equation}
and  hereafter  the  notation
\begin{equation}
\dfrac{\beta_i+\gamma_i \hr_i}{\hat{x}_i^2}=\dfrac{1}{\hat{x}_i^2}\big(\beta_i+\gamma_i \hr_i \big)=\big(\beta_i+\gamma_i \hr_i \big)\dfrac{1}{\hat{x}_i^2} 
\label{notation1}
\end{equation}
is assumed,  which is consistent since $ \big[\hr_i,\hx^2_i \big]=0$.

The first objective of the paper is achieved by the following statement.
\begin{prop}
\label{prop1}
(i) For any choice of the function $F$, the Hamiltonian $\hat{H} $ (\ref{ham}) is endowed with  $2N-3$ ``universal'' left and right quantum integrals of Dunkl type given by 
\begin{equation}
\begin{split}
\hat{C}^{[m]}& =\quad\   \sum_{1 \leq i<j }^m \quad \,\left(\hl_{ij}^2+\big(\beta_i + \gamma_i \hr_i \big)\frac{\hx_j^2}{\hx_i^2}+\big(\beta_j + \gamma_j \hr_j \big)\frac{\hx_i^2}{\hx_j^2} \right)\,  +\quad \sum_{i=1}^m \big(\beta_i+\gamma_i \hr_i \big)\\[2pt]
 \hat{C}_{[m]}&= \sum_{N-m+1 \leq i<j }^N \left(\hl_{ij}^2+\big(\beta_i + \gamma_i \hr_i \big)\frac{\hx_j^2}{\hx_i^2}+\big(\beta_j + \gamma_j \hr_j \big)\frac{\hx_i^2}{\hx_j^2} \right)\, + \!\sum_{i=N-m+1}^N\!\!\big(\beta_i+\gamma_i \hr_i \big)  
 \end{split}
\label{left1}
\end{equation}
where $m=2, \dots, N$ and such that  
\begin{equation}
\hat{C}^{[N]} = \hat{C}_{[N]}  =  \sum_{1 \leq i<j }^N \left(\hl_{ij}^2+\big(\beta_i + \gamma_i \hr_i \big)\frac{\hx_j^2}{\hx_i^2}+\big(\beta_j + \gamma_j \hr_j \big)\frac{\hx_i^2}{\hx_j^2} \right)\,+\,\sum_{i=1}^N  \big(\beta_i+\gamma_i \hr_i ) \, .
\label{CN}
\end{equation}
(ii) Each set $\big\{\hat{H},\hat{C}^{[2]},\dots, \hat{C}^{[N]}\big\}$ and $\big\{\hat{H},\hat{C}_{[2]},\dots, \hat{C}_{[N]}\big\}$ is formed by $N$ algebraically independent commuting operators:
\begin{equation}
\left[\hat{C}^{[m]}, \hat{H}\right]=\left[\hat{C}_{[m]}, \hat{H}\right]=0 \, , \quad \left[\hat{C}^{[m]}, \hat{C}^{[m']}\right]=\left[\hat{C}_{[m]}, \hat{C}_{[m']}\right]=0 \, ,
\label{eq:comm}
\end{equation}
where $m,m'=2, \dots, N$.\\
(iii) The set $\big\{\hat{H},\hat{C}^{[2]},\dots, \hat{C}^{[N]} , \hat{C}_{[2]},\dots, \hat{C}_{[N-1]}\big\}$ consists of $2N-2$  algebraically independent operators.
 \end{prop}

Before presenting the technical details that allow this result to be established, some observations are in order. 
 \begin{rem}
 Assuming that $\hat{H}$ is different from the operator $ \hat{C}^{[N]} = \hat{C}_{[N]} $, the 
  whole family (\ref{ham}) is not only integrable, due to the commutators (\ref{eq:comm}), but it
   turns out to be QMS, which means that for any choice of the function $F$ and any values of the parameters $\{\beta_i, \mu_i, \gamma_i\}_{i=1}^N$ the resulting Hamiltonian is always endowed with the same set of $2N-3$  ``universal" quantum integrals  \eqref{left1}.  The term  ``universal" comes from the nomenclature introduced in \cite{BH2007} for the constants of motion of QMS classical systems provided by the $\mathfrak{sl}(2,\mathbb{R})$  coalgebra and   used subsequently in \cite{Latini2019} to describe the quadratic algebras they generate.    
\end{rem}
  
\begin{rem}
 The terminology ``left" and ``right" for the quantum operators  \eqref{left1} refers to the sequence
 of the indices in the summations which ensures the property of algebraical independence of the 
 universal  $2N-3$  quantum integrals.     In fact, they ultimately come from  the so-called left and right Casimirs of the $\mathfrak{sl}(2,\mathbb{R})$ coalgebra~\cite{BHMR2004,BalBlaHerMusRag2009} adapted to the Dunkl framework, as we will show below.
  \end{rem}

  \begin{rem}
  \label{rem4}
 For $\{\beta_i, \mu_i,  \gamma_i\}_{i=1}^N =0$, we recover the $N$D family of Hamiltonians $\hat{H} = F\bigl(\hbp^2, \hbx^2,\hbx \cdot \hbp-\imath \hbar N/2\bigl)$ \cite{BEHRR2011}, which is characterized by an $\mathfrak{so}(N)$ symmetry. In this case, the quantum integrals emerge as sums of the squares of the angular momentum components, as they correspond to the quadratic Casimirs of certain rotational subalgebras $\mathfrak{so}(m) \subseteq \mathfrak{so}(N)$. Such a spherical symmetry can be generalized with the appearence of non-central terms in the Hamiltonian, considering $\{\beta_i\}_{i=1}^N \neq 0$, as discussed for example in \cite{Latini2019}. In turn, if $\{\mu_i\}_{i=1}^N\neq0$, we get an extension involving Dunkl operators. Finally, if we also take $\{\gamma_i\}_{i=1}^N\neq0$ we get a (full) Dunkl extension of the Hamiltonian involving non-central terms that also accomodate reflection operators.
 From this point of view, we can say that the set  \eqref{left1}  consists of  ``universal" quantum integrals of  Dunkl type  as they   cover all  the  particular situations mentioned above. 
 \end{rem}

\begin{rem}
The left and right integrals (\ref{left1})  can also be  rewritten as ($m=2, \dots, N$):
\begin{equation}
\begin{split}
&\hat{C}^{[m]}  = \quad \sum_{1 \leq i<j }^m \quad \,\,\, \hl_{ij}^2 \,+ \left(\sum_{i=1}^m \hx_i^2\right) \left(\sum_{i=1}^m \frac{\beta_i + \gamma_i \hr_i}{\hx_i^2}\right)  \\[2pt]
&\hat{C}_{[m]}  = \sum_{N-m+1 \leq i<j }^N  \,\hl_{ij}^2 \,+ \left(\sum_{i=N-m+1}^N \hx_i^2\right) \left(\sum_{i=N-m+1}^N \frac{\beta_i + \gamma_i \hr_i}{\hx_i^2}\right) \,  .
 \end{split}
\label{eq:7}
\end{equation}
\end{rem}

\begin{rem}
MS property only  arises  for very specific choices of the function $F$.    We will present explicit MS systems of Dunkl type in  Sections \ref{sec4} and  \ref{sec5}.   
\end{rem}

  The results by Proposition~\ref{prop1} arise as a direct consequence of the coalgebra symmetry method \cite{BalCorRag1996, BalRag1998, BalBlaHerMusRag2009}:  the Hamiltonian \eqref{ham}, for any suitable choice of the function $F$, is endowed with an $\mathfrak{sl}(2,\mathbb{R})$ coalgebra symmetry through a specific differential-difference realization. In this way, the quantum integrals \eqref{left1}  emerge as the images, under such a realization, of the left and right partial Casimirs  of $\mathfrak{sl}(2,\mathbb{R})$. In what follows, we make this point explicit by presenting a new differential-difference realization of the Lie algebra $\mathfrak{sl}(2,\mathbb{R})$ with the aim of bridging the gap between the coalgebra symmetry formalism and Dunkl superintegrable systems.
 
  Let us consider the $\mathfrak{sl}(2, \mathbb{R}) \cong \mathfrak{su}(1,1)$ Lie algebra whose generators $J_\al$ ($\al = \pm, 3$) satisfy the commutation relations:
\begin{equation}
[J_-, J_+] = 4 \imath \hbar J_3 \, , \quad [J_3, J_\pm] = \pm 2 \imath  \hbar J_\pm \, .
\label{sl2}
\end{equation}
We also introduce the (primitive or non-deformed) coproduct $\Delta: \mathfrak{sl}(2,\mathbb{R}) \to \mathfrak{sl}(2,\mathbb{R}) \otimes \mathfrak{sl}(2,\mathbb{R})$ such as:
\begin{equation}
\Delta(1) = 1 \otimes 1 \, , \quad \Delta(J_\al) = J_\al \otimes 1 + 1 \otimes J_\al \, ,
\label{cop}
\end{equation}
which is a Lie algebra homomorphism i.e., 
\begin{equation}
 \Delta (X \cdot Y)=\Delta (X)\cdot \Delta (Y) \, , \quad \forall\, X,Y \in \mathfrak{sl}(2,\mathbb{R}) \quad \Rightarrow\quad   \Delta\big(\big[J_\al,J_{\al'} \big] \big)=\big[\Delta(J_\al),\Delta(J_{\al' })\big]  \,  ,\quad \al,\al'=\pm, 3
\label{eq:homprop}
\end{equation}
 and fulfils the coassociativity condition
\begin{equation}
   (\text{id} \otimes \Delta ) \circ \Delta =(\Delta \otimes \text{id}) \circ \Delta \, .
\label{eq:copmapext}
\end{equation}
 Notice that the latter requirement is defined on the 3D tensor product space $\mathfrak{sl}(2,\mathbb{R}) \otimes \mathfrak{sl}(2,\mathbb{R}) \otimes \mathfrak{sl}(2,\mathbb{R})$.
Recall also that the pair $\big(\mathfrak{sl}(2, \mathbb{R}) ,\Delta\big)$ determines a Lie coalgebra, which can be  extended to the well-known  Hopf algebra structure that exists for any Lie algebra~\cite{CP94,Abe2004}.

The Lie algebra  is endowed with the following ``seed" Casimir: 
\begin{equation}
C=C(J_+,J_-,J_3) =\frac{1}{2} (J_+J_-+J_-J_+)-J_3^2 \, ,
\label{eq:seed}
\end{equation}
belonging to its universal enveloping algebra. Clearly, by definition, for this quadratic element we have:
\begin{equation}
[C,J_\al]=0 \, , \quad \al = \pm, 3\, .
\label{cas1}
\end{equation}
  Let us now introduce the following 1D differential-difference realization:
\begin{equation}
\left\{\begin{aligned}
\hat{J}_+^{[1]}&:= D(J_+)=\hpi_1^2+\dfrac{\beta_1+\gamma_1 \hr_1}{\hat{x}_1^2} \\
\hat{J}_3^{[1]}&:=D(J_3) \hskip 0.075cm=\hx_1 \hpi_1- \imath \hbar \big(\tfrac 12+\mu_1 \hr_1 \big) \\[2pt] 
\hat{J}_-^{[1]}&:= D(J_-)=\hx_1^2 \, 
\end{aligned}\right. 
\label{eq:sl2rep}
\end{equation}
 where we have used the notation (\ref{notation1}), for which the following commutation relations are verified:
\begin{equation}
\left[\hat{J}_-^{[1]},\hat{J}_+^{[1]}\right]= 4 \imath \hbar \hat{J}_3^{[1]}\, , \quad \left[\hat{J}_3^{[1]},\hat{J}_\pm^{[1]}\right]= \pm 2 \imath \hbar \hat{J}_\pm^{[1]} \, .
\label{eq:11PBssl2quantum}
\end{equation}
 Note that in the realization (\ref{eq:sl2rep})  both the Dunkl momentum operator $\hpi_1$ and a reflection operator $\hr_1$ appear, the latter connected to the parameters $\mu_1$ and $\gamma_1$, respectively.

 We find it meaningful to provide the explicit expression for the realization in terms of the standard momentum $\hat{p}_1$  in order to show how the generators are extended with respect to the ordinary quantum case.   By expanding the Dunkl momentum operator (\ref{dunklmom}), we obtain:
\begin{equation}
\left\{\begin{aligned}
 \hat{J}_+^{[1]}&=\hp_1^2-2 \imath \hbar \dfrac{\mu_1}{\hat{x}_1}\hat{p}_1+\dfrac{1}{\hat{x}_1^2}	\left( \big(\beta_1+\hbar^2 \mu_1 \big)+\big(\gamma_1-\hbar^2 \mu_1\big) \hr_1\right) \\ 
\hat{J}_3^{[1]}&=\hat{x}_1 \hat{p}_1-\imath \hbar\big(\tfrac 12+\mu_1\big) \\[2pt]
\hat{J}_-^{[1]}&=\hx_1^2 \, . 
\end{aligned}\right. 
\label{eq:sl2repexp}
\end{equation}
Thus, it is readily verified that the standard quantum realization (see for example \cite{Latini2019}) is obtained by setting $\mu_1=0$ and $\gamma_1=0$, as expected.
Concerning the expression of the Casimir, under the above realization, it turns out to be
\begin{equation}
 \hat{C}^{[1]} :=\hat{C}\big(\hat{J}_+^{[1]},\hat{J}_-^{[1]},\hat{J}_3^{[1]}\big) = \frac{1}{2}\left(\hat{J}_+^{[1]} \hat{J}_-^{[1]}+\hat{J}_-^{[1]} \hat{J}_+^{[1]}\right)-\left(\hat{J}_3^{[1]}\right)^2 = (\imath \hbar)^2\left( \frac 34-\mu_1\big(\mu_1-\hr_1\big)\right)+ \beta_1+\gamma_1 \hr_1  \, .
\label{cas1Dquantums}
\end{equation}

\begin{rem}
\label{rem7}
 It is worth observing that  this result is quite different from the standard case (without reflections) corresponding to set $\mu_1=\gamma_1=0$, such that $\hat{C}^{[1]}$ reduces to the constant: $-\frac 34 \hbar^2  + \beta_1$. Therefore, when reflections are considered, $\hat{C}^{[1]}$ is no longer a constant but   is determined by the reflection operator $\hr_1$.  In this respect, we recall that in the classical picture  with the Lie-Poisson $\mathfrak{sl}(2,\mathbb{R}) $ coalgebra, the realization simply gives    $ {C}^{[1]}= \beta_1$, which means that the sign of    $ \beta_1$ labels the representation of  $\mathfrak{sl}(2,\mathbb{R}) $. To be precise,  $ \beta_1$ characterizes the  three types  of symplectic submanifolds corresponding to the surfaces with constant value $\beta_1$, hence the two (non-equivalent) representations with $\beta_1\ne 0$  provide a non-central potential through the Poisson generator ${J}_+^{[1]}=p_1^2 + {\beta_1}/{ {x}_1^2}$.
\end{rem}

Clearly, since $ \hat{C}^{[1]} $ represents a Casimir element of the Lie algebra, it commutes with all the three quantum generators not only abstractly (\ref{cas1}), but also in the given realization. Such commutation will be preserved if one considers suitable functions of the generators, that we interpret as the Hamiltonian, i.e.:
\begin{equation}
 \hat{H} =F\left(\hat{J}_+^{[1]},\hat{J}_-^{[1]},\hat{J}_3^{[1]}\right)\, .
 \label{eq:hamgen1d}
 \end{equation}
However, we observe that, at the 1D level, the quantum constant obtained through the Casimir operator is trivial, as it does not explicitly depend on the momenta. This situation dramatically changes when higher-dimensional tensor product spaces and, consequently, higher-dimensional realizations of the Lie algebra are considered. Let us explore this in detail.

  \noindent The generalization to $N$D is obtained by considering the $N$-th coproduct $\Delta^{[N]}: \mathfrak{sl}(2,\mathbb{R}) \to  \mathfrak{sl}(2,\mathbb{R}) \otimes\dots^{N)} \otimes \mathfrak{sl}(2,\mathbb{R})$, where from now on we will make use of the notation:
\begin{equation}
a \otimes \dots^{\kk)} \otimes a \equiv \overbrace{a \otimes a \otimes \dots \otimes  a}^{\kk \,\,\text{times}} \, .
\label{notation}
\end{equation}
 In particular, within this approach, the coproduct map is extended recursively by considering a generalization of the coassociativity property (\ref{eq:copmapext}). At this level, in fact, it allows us to define the new coproduct $\Delta^{[3]}: \mathfrak{sl}(2,\mathbb{R}) \to \mathfrak{sl}(2,\mathbb{R})  \otimes \mathfrak{sl}(2,\mathbb{R})  \otimes \mathfrak{sl}(2,\mathbb{R})$ in two ways as:
 \begin{equation}
 	\Delta^{[3]}: = \big(\text{id} \otimes \Delta^{[2]} \big ) \circ \Delta^{[2]} = \big(\Delta^{[2]} \otimes \text{id} \big) \circ \Delta^{[2]}  \, ,
 	\label{eq:copmapext3}
 \end{equation}
where we have introduced the notation $\Delta^{[2]}:=\Delta$. The general $N$-th coproduct is defined recursively in the form
 \begin{equation}
	\Delta^{[N]}:= \big(\text{id} \otimes \dots^{N-2)} \otimes   \text{id} \otimes \Delta^{[2]} \big) \circ \Delta^{[N-1]} =\big(\Delta^{[2]} \otimes  \text{id} \otimes \dots^{N-2)}\otimes   \text{id} \big) \circ \Delta^{[N-1]} \, ,
	\label{eq:copmapextN}
\end{equation}
and it allows to extend the generators from $\mathfrak{sl}(2,\mathbb{R})$ to $\mathfrak{sl}(2,\mathbb{R}) \otimes\dots^{N)} \otimes \mathfrak{sl}(2,\mathbb{R})$ as
\begin{equation}
J_+^{[N]}:=\Delta^{[N]}(J_+) \, , \quad J_-^{[N]}:=\Delta^{[N]}(J_-) \, , \quad J_3^{[N]}:=\Delta^{[N]}(J_3) \, .
\label{eq:NDgen}
\end{equation}
Their explicit expression on the tensor product space $\mathfrak{sl}(2,\mathbb{R}) \otimes\dots^{N)} \otimes \mathfrak{sl}(2,\mathbb{R})$ reads
\begin{equation}
J_\al^{[N]}=J_\al \otimes 1 \otimes \dots^{N-1)} \otimes 1 + 1 \otimes J_\al \otimes 1 \otimes \dots^{N-2)} \otimes 1 + \dots +1 \otimes  \dots^{N-1)} \otimes 1 \otimes J_\al\,  \quad (\al=\pm,3) \, .
\label{eq:expNDgen}
\end{equation}
Similarly to the initial coproduct $\Delta$  (\ref{eq:homprop}), the $N$-th coproduct is also an algebra homomorphism between $\mathfrak{sl}(2,\mathbb{R})$ and   $\mathfrak{sl}(2,\mathbb{R}) \otimes\dots^{N)} \otimes \mathfrak{sl}(2,\mathbb{R})$,   and we have:
\begin{equation}
 \Delta^{[N]}(X \cdot Y)=\Delta^{[N]}(X)\cdot \Delta^{[N]}(Y) \, , \quad \forall\, X,Y \in \mathfrak{sl}(2,\mathbb{R}) \, . 
\label{eq:homprop2}
\end{equation}
If we restrict our attention to the generators, this implies that
 \begin{equation}
\begin{split}
\left[\Delta^{[N]}(J_-),\Delta^{[N]}(J_+)\right]_{\mathfrak{sl}(2,\mathbb{R}) \otimes\dots^{N)} \otimes \mathfrak{sl}(2,\mathbb{R})}&=\Delta^{[N]}\left([J_-,J_+]_{\mathfrak{sl}(2,\mathbb{R})}\right)=4\imath \hbar\Delta^{[N]}(J_3)\\
\left[\Delta^{[N]}(J_3),\Delta^{[N]}(J_\pm)\right]_{\mathfrak{sl}(2,\mathbb{R}) \otimes\dots^{N)} \otimes \mathfrak{sl}(2,\mathbb{R})}&=\Delta^{[N]}\left([J_3,J_\pm]_{\mathfrak{sl}(2,\mathbb{R})}\right)=\pm2\imath \hbar\Delta^{[N]}(J_\pm) \, .
 \end{split}
\end{equation}
As a direct consequence, the images of these extended generators under the   tensor product of the  differential-difference realization (\ref{eq:sl2rep}) in any dimension $N$ turn out to be
 \begin{equation}
\left\{\begin{aligned}
\hat{J}_+^{ [N]} &:= (D \otimes\dots^{N)} \otimes D )\Delta^{[N]}(J_+)=\mathbf{\hbpi}^2+ \sum_{i=1}^N \left(\frac{\beta_i+\gamma_i \hr_i}{\hx_i^2}\right) \\
\hat{J}_3^{ [N]}&:= (D  \otimes\dots^{N)} \otimes D)  \Delta^{[N]}(J_3)\hskip 0.075cm= \mathbf{\hbx} \cdot \mathbf{\hbpi} -\imath \hbar \left(\frac N2+\sum_{i=1}^N \mu_i \hr_i \right)\\
\hat{J}_-^{ [N]}& := (D  \otimes\dots^{{N)}} \otimes D ) \Delta^{[N]}(J_-) = \hbx^2 \,  
\end{aligned}\right. 
\label{NDrepquantum}
\end{equation}
which automatically satisfy the same defining commutation rules of  $\mathfrak{sl}(2, \mathbb{R})$ (\ref{sl2}):
\begin{equation}
\left [\hat{J}_-^{ [N]} ,\hat{J}_+^{ [N]}\right ]= 4 \imath \hbar \hat{J}_3^{ [N]} \, ,\quad  \, \left[\hat{J}_3^{[N]},\hat{J}_{\pm}^{[N]}\right]= \pm 2 \imath\hbar \hat{J}_{\pm}^{[N]}\, .
\label{eq:18PBssl2quantum}
\end{equation}

\begin{rem}
The realization (\ref{NDrepquantum}) can also be expressed in terms  of the standard momentum operator $\hbp$, as in the 1D case (\ref{eq:sl2repexp}); in particular, we have that
\begin{equation}
\hat{J}_3^{ [N]}=\mathbf{\hbx} \cdot \mathbf{\hbpi} -\imath \hbar \left(\frac N2+\sum_{i=1}^N \mu_i \hr_i \right) = \hbx \cdot \hbp-\imath \hbar \left(\frac N2+\sum_{i=1}^N \mu_i\right)\, .
\end{equation}
\end{rem}

\noindent
Now, here is the crucial point behind this  algebraic approach: it turns out that the extended generators commute with the $2N - 3$ \textquotedblleft left\textquotedblright and \textquotedblleft right\textquotedblright partial Casimirs of the coalgebra. For $m=2, \dots, N$, these elements read:
\begin{equation}
C^{[m]}=\Delta^{[m]}(C) \otimes 1 \otimes \dots^{N-m)} \otimes 1\, , \quad C_{[m]}= 1 \otimes \dots^{N-m)} \otimes 1 \otimes \Delta^{[m]}(C) \, ,
\label{LeftRight}
\end{equation}
where $\Delta^{[m]}$  is the $m$-th coproduct. Notice that $C^{[N]} =C_{[N]}$. So, they arise as a  \textquotedblleft left\textquotedblright and  \textquotedblleft right\textquotedblright embedding of the $m$-th coproducts into the tensor product space $\mathfrak{sl}(2,\mathbb{R}) \otimes\dots^{N)} \otimes \mathfrak{sl}(2,\mathbb{R})$. 
\begin{rem}
The previous expressions also accomodate the case $m=1$. In this case, we have:
\begin{equation}
C^{[1]}=\Delta^{[1]}(C) \otimes 1 \otimes \dots^{N-1)} \otimes 1\, , \quad C_{[1]}= 1 \otimes \dots^{N-1)} \otimes 1 \otimes \Delta^{[1]}(C) \, ,
\label{LeftRightm1}
\end{equation}
where $\Delta^{[1]}=\text{id}$. This means that this case collapses to:
\begin{equation}
	C^{[1]}=C \otimes 1 \otimes \dots^{N-1)} \otimes 1\, , \quad C_{[1]}= 1 \otimes \dots^{N-1)} \otimes 1 \otimes C \, .
	\label{LeftRightm1c}
\end{equation}
However, we have not considered it further, as it does not yield any nontrivial results from the perspective of quantum integrals. This point will become clearer in the discussion below.
\end{rem}

 The set of left Casimirs $C^{[m]}$ commute with the generators \eqref{eq:NDgen} in the tensor product space  $\mathfrak{sl}(2,\mathbb{R}) \otimes\dots^{N)} \otimes \mathfrak{sl}(2,\mathbb{R})$ and, furthermore, they are in involution among themselves \cite{BalRag1998}. The same applies to the other set of right Casimirs $C_{[m]}$, the latter being considered for the first time in \cite{BHMR2004}. The commutation of the partial Casimirs with the generators can be extended to suitable (smooth or formal power series) functions of the above generators.
 
As a consequence, applying the differential-difference realization (\ref{NDrepquantum}),  replacing the index $N$ by $m$, to the expressions (\ref{LeftRight}) with   $C$ given by (\ref{eq:seed}), we obtain the images of these partial Casimirs   leading to the following (universal) quantum integrals:

 \begin{equation}
\begin{split}
\hat{C}^{[m]} =   &\sum_{1 \leq i<j }^m  \left(\hl_{ij}^2+\big(\beta_i + \gamma_i \hr_i \big)\frac{\hx_j^2}{\hx_i^2}+\big(\beta_j + \gamma_j \hr_j \big)\frac{\hx_i^2}{\hx_j^2} \right)\,+\,\sum_{i=1}^m \big(\beta_i+\gamma_i \hr_i \big)   \\
& +\hbar^2 \left(\frac{m(m-4)}{4} +\sum_{i=1}^m \mu_i \big(\mu_i +(m-2) \hr_i \big)+2 \sum_{1 \leq i<j }^m \mu_i \mu_j \hr_i\hr_j \right) \\[2pt]
 \hat{C}_{[m]}  = & \sum_{N-m+1 \leq i<j }^N \left(\hl_{ij}^2+\big(\beta_i + \gamma_i \hr_i \big)\frac{\hx_j^2}{\hx_i^2}+\big(\beta_j + \gamma_j \hr_j \big)\frac{\hx_i^2}{\hx_j^2} \right)\,+\,\sum_{i=1}^N  \big(\beta_i+\gamma_i \hr_i \big)   \\[2pt]
& +\hbar^2 \left(\frac{m(m-4)}{4} +\sum_{i = N-m+1}^N \mu_i \big(\mu_i +(m-2) \hr_i \big)+2 \sum_{N-m+1 \leq i <j}^N \mu_i \mu_j \hr_i\hr_j\right)\, ,
  \end{split}
\label{fullLeftRight}
\end{equation}
with $m=2, \dots, N$ and $\hat{C}^{[N]}=\hat{C}_{[N]}$.

Notice that if we take $m=1$ at a fixed $N$ in \eqref{fullLeftRight}, we obtain:
 \begin{equation}
\begin{split}
 \hat{C}^{[1]} &= \beta_1+\gamma_1 \hr_1 -  \hbar^2\left(\frac 34-\mu_1\big(\mu_1-\hr_1\big)\right)  \\[2pt]
   \hat{C}_{[1]} &=\beta_N+\gamma_N \hr_N -  \hbar^2\left(\frac 34-\mu_N\big(\mu_N-\hr_N\big)\right)\, ,
 \end{split}
\label{meq1}
\end{equation}
so that the former is just (\ref{cas1Dquantums}). We discard these quantum integrals as they are not polynomial in the momenta.

  The resulting quantum integrals (\ref{fullLeftRight})  appear naturally as the sum of two distinct terms:
\begin{equation}
\hat{C}^{[m]}=\hat{C}^{[m]}_1+\hbar^2\, \hat{C}^{[m]}_2 \, ,\quad \hat{C}_{[m]}=\hat{C}_{[m],1}+\hbar^2\, \hat{C}_{[m],2} \, ,
\label{eq:coalref}
\end{equation}
where the former involves the sum of the square of  Dunkl angular momentum operators (\ref{eq:angmomdunkl}), whereas the latter  depends  only on reflection operators. It turns out that these operators commute separately with the coalgebra generators in the given realization, i.e.: 
\begin{equation}
\left[\hat{C}^{[m]}_{\ii}, \hat{J}_{\al}^{[N]}\right]=\left[\hat{C}_{[m],\ii}, \hat{J}_{\al}^{[N]}\right]=0 \, , \quad m=2, \dots, N \, , \quad \ii=1,2 \, ,\quad \al= \pm,3  \,.
\end{equation}
Consequently, they also commute with the Hamiltonian  \eqref{ham} (which is here intended to be a suitable function of the aforementioned generators), i.e.:
\begin{equation}
 \hat{H}=F\left(\hat{J}_+^{[N]},\hat{J}_-^{[N]},\hat{J}_3^{[N]}\right)  \, .
 \label{eq:ham}
\end{equation}
For these quantum operators, it also holds: 
\begin{equation}
\left[\hat{C}^{[m]}_{\ii}, \hat{C}^{[m']}_\ii\right]=\left[\hat{C}_{[m],\ii}, \hat{C}_{[m'],\ii}\right]=0 \, , \quad  m,m'=2, \dots, N \, , \quad \ii=1,2\,,
\label{commutconstants}
\end{equation}
as can be checked by direct computations. This allows us to extrapolate the (genuine) quantum integrals \eqref{left1} (or, equivalently, \eqref{eq:7}), which are all quadratic in the momenta through the square of the Dunkl angular momentum operators (\ref{eq:angmomdunkl}). These integrals are labelled here with the subscript “1” (which is  omitted in \eqref{left1}), and lead to the desired result stated in Proposition~\ref{prop1}.


\section{Quasi-maximally superintegrable systems of Dunkl type with central potentials}
\label{sec3}
\noindent 
A natural application of Proposition~\ref{prop1} is the construction of curved Dunkl type systems with central potentials along with their generalizations via the superposition with $N$ arbitrary non-central potentials.
With this aim, let us first consider the Euclidean space $\bE^N$ with the usual Cartesian coordinates $\bx$.
In the  standard simplest case with $\{\beta_i, \mu_i,  \gamma_i\}_{i=1}^N =0$,  is well-known that the quantum Euclidean Hamiltonian with an arbitrary central potential is always QMS, but generically non-MS,  namely 
\begin{equation}
\hat{H} = \frac{\hbp^2}{2}+ V(|\hbx|) =-\frac{\hbar^2}{2}\Delta +V(|\bx|)\, , \quad \Delta = \sum_{i=1}^N \partial_i^2 \,  ,
\label{eq:Euclidean1}
\end{equation}
such that hereafter $\Delta$ denotes the usual Laplacian.
This result can  be retrieved  straightforwardly from the $\mathfrak{sl}(2,\mathbb{R}) $ coalgebra symmetry. By taking into account that  the realization (\ref{NDrepquantum}) reduces to 
\begin{equation}
\hat{J}_+  =\mathbf{\hbp}^2\, , \quad \hat{J}_-  =\mathbf{\hbx}^2\, , \quad \hat{J}_3   = \mathbf{\hbx} \cdot \mathbf{\hbp}-\imath \hbar \frac N2 \, ,
\label{repr0}
\end{equation}
where from now on we omit the index ``$[N]$"  as we always consider $N$D systems unless otherwise stated.  The Hamiltonian can be expressed in terms of the function (\ref{eq:ham}) as
\be
\hat{H} = \frac {\hat{J}_+}2  +  V\!\left(\!\sqrt{\hat{J}_-} \,\right)\, .
\label{HamC}
\ee
It is immediate to derive the corresponding common set of $2N-3$ quantum integrals from (\ref{left1}), which are sums of the square of certain standard angular momentum operators (\ref{standardL}) as commented in Remark~\ref{rem4}.
The paradigmatic MS systems  \cite{Bertrand1873} correspond to the isotropic harmonic oscillator and KC potential, whose additional ``hidden" quantum symmetries are quadratic in the momenta and given, in this order, by the  Demkov-Fradkin  tensor \cite{D1959, F1965} and the  LRL $N$-vector \cite{G1975, G1976, GPS2002}; these potentials are directly identified in  (\ref{HamC})   in the form 
 \begin{equation}
\begin{split}
V_{\textsf{HO}}&= \frac{ \omega^2}2  {\hat{J}_-}  =\frac  {\omega^2}{2} \hbx^2  \, \quad (\omega \in \mathbb{R})  \\
V_{\textsf{KC}}&=-\frac{k}{ {\hat J_-^{1/2}}}  =-\frac{k}{|\hbx|}   \,  \quad (k \in \mathbb{R})\,   .
  \end{split}
  \label{HOKC}
\end{equation}
Then, up to $N$ arbitrary non-central $\beta_i$-potentials arise by keeping the same Hamiltonian function (\ref{HamC}), but now
taking  the realization (\ref{NDrepquantum}) with non-zero $\beta_i$'s yielding
\begin{equation}
\hat{H} = \frac{\hbp^2}{2}+ \sum_{i=1}^N \frac{\beta_i}{2\hx_i^2}+ V(|\hbx|) =-\frac{\hbar^2}{2}\Delta+   \sum_{i=1}^N \frac{\beta_i}{2 x_i^2}
+V(|\bx|)\, ,
\label{eq:Euclidean2}
\end{equation}
providing the so-called generalized Hamiltonian, which is also QMS (see~\cite{Evans90b}  for classical systems with $N=3$).  The appearance of each $\beta_i$-potential comes from the free Hamiltonian $\hat \TT= {\hat{J}_+}/2$ and is a direct consequence of the chosen representation for the $i$-th copy of $\mathfrak{sl}(2,\mathbb{R})$ within the tensor product
${\mathfrak{sl}(2,\mathbb{R}) \otimes\dots^{N)} \otimes \mathfrak{sl}(2,\mathbb{R})}$, since the realization of  the corresponding Casimir, obtained by (\ref{cas1Dquantums}) substituting the index 1 by $i$, reads
\begin{equation}
 \hat{C}^{[i]} = \frac{1}{2}\left(\hat{J}_+^{[i]} \hat{J}_-^{[i]}+\hat{J}_-^{[i]} \hat{J}_+^{[i]}\right)-\left(\hat{J}_3^{[i]}\right)^2 =-  \frac 34  \hbar^2 + \beta_i  \, .
\label{cas1Dquantums2}
\end{equation}
As pointed out in Remark~\ref{rem7}, a positive, negative or zero value of $\beta_i$ determines the specific representation of the $i$-th copy of $\mathfrak{sl}(2,\mathbb{R})$. 
Such non-central potentials are known as Rosochatius or Winternitz terms and their role 
 is well-known  in classical mechanics: each of them gives rise to a ``centrifugal" barrier which limits the trajectory of the particle; hence, they are also called  centrifugal terms. In this paper, for a given Hamiltonian with a central potential $V(|\hbx|)$,  the superposition with all     $N$ $\beta_i$-terms is referred to as the   generalized Hamiltonian.  Specifically, the generalized oscillator system of $V_{\textsf{HO}}$ (\ref{HOKC}) is the so-called Smorodinsky-Winternitz system~\cite{WSUF65}  (see also   \cite{Evans90,Evans91,GPS1995}), which is endowed with additional quantum integrals that reduce to the diagonal components of the Demkov-Fradkin  tensor  when the non-central terms are discarded (they are essentially the one-dimensional Hamiltonians arising from separability in Cartesian coordinates). Analogously, the generalized KC system of $V_{\textsf{KC}}$ (\ref{HOKC}) can be constructed but, however, there only exist
 LRL type quantum integrals quadratic in the momenta whenever, at least, one $\beta_i$-term vanishes; in this case, the resulting Hamiltonian determines the so-called quasi-generalized KC system \cite{Evans90b,RW2002,KWMP2002}. The proper generalized KC system is endowed with $N$ LRL type integrals quartic in the momenta \cite{VE2008,KC2009,TD2011}.

  Similarly, in this standard quantum setting, an extension of the Hamiltonian (\ref{eq:Euclidean2}) is again obtained  by  the realization (\ref{NDrepquantum}) by taking up to $N$ non-zero parameters $\gamma_i$ entailing the operators $\hr_i$, which can be introduced regardless of the $\beta_i$-potentials, while keeping the same formal Hamiltonian (\ref{HamC}),   thus arriving at
 \begin{equation}
\hat{H} = \frac{\hbp^2}{2}+ \sum_{i=1}^N \frac{\beta_i+\gamma_i \hr_i}{2\hx_i^2}+ V(|\hbx|) =-\frac{\hbar^2}{2}\Delta+   \sum_{i=1}^N \frac{\beta_i+\gamma_i \hr_i}{2x_i^2}+V(|\bx|) \, ,
\label{eq:Euclidean3}
\end{equation}
so that the Rosochatius-Winternitz terms become centrifugal-reflection non-central potentials. Therefore, we are now dealing with a differential-difference realization that manifests itself in the Casimir of the 
 $i$-th copy of $\mathfrak{sl}(2,\mathbb{R})$ since it is also determined by a reflection operator  $\hr_i$ (as discussed in Remark~\ref{rem7}):
 \begin{equation}
 \hat{C}^{[i]} =  -  \frac 34  \hbar^2 + \beta_i +\gamma_i\hr_i\,  .
\label{cas1Dquantums3}
\end{equation}
In this respect, recall that reflection operators have already been used in the standard quantum context (see, e.g., \cite{PVZ2011,MP2024} and references therein).

 The  Hamiltonian (\ref{eq:Euclidean3}),  formed by the superposition of a central potential with  $N$  $\beta_i$-centrifugal and $N$ $\gamma_i$-reflection  non-central potentials on $\bE^N$, is finally extended to the Dunkl framework through the most general differential-difference realization (\ref{NDrepquantum}), keeping once more the same Hamiltonian function (\ref{HamC}), explicitly 
  \begin{equation}
\hat{H} = \frac {\hat{J}_+}2  +  V\!\left(\!\sqrt{\hat{J}_-} \,\right)=\frac{\hbpi^2}{2}+ \sum_{i=1}^N \frac{\beta_i+\gamma_i \hr_i}{2\hx_i^2}+ V(|\hbx|) =-\frac{\hbar^2}{2}\Delta_\D+   \sum_{i=1}^N \frac{\beta_i+\gamma_i \hr_i}{2x_i^2}+V(|\bx|)
\, .
\label{eq:Euclideanjaddterms}
\end{equation}
In this way, we obtain a particular family of QMS generalized systems of  Dunkl type  on $\bE^N$ for an arbitrary central potential  and parameters $\{\beta_i , \gamma_i \}_{i=1}^N \neq 0$, which is always endowed with  $2N-3$ quantum integrals by Proposition~\ref{prop1}.
Let us note that the Dunkl counterpart of the $\mathfrak{sl}(2, \mathbb{R})$ realization (\ref{repr0}),  
with  $\{\beta_i , \gamma_i \}_{i=1}^N = 0$ in \eqref{NDrepquantum}, is given by
\begin{equation}
\hat{J}_+  =\mathbf{\hbpi}^2 \, , \quad \hat{J}_-  = \hbx^2\, , \quad \hat{J}_3   = \mathbf{\hbx} \cdot \mathbf{\hbpi} -\imath \hbar \left(\frac{N}2+\sum_{i=1}^N \mu_i \hr_i \right) \, ,
\label{NDrepquantumdunkl}
\end{equation}
through which we find that
\begin{equation}
\hat{H} = \frac{\hat{J}_+}{2}+ V\left(\sqrt{\hat{J}_-}\right) =-\frac{\hbar^2}{2}\Delta_\D+V(|\bx|)\, .
\label{eq:Euclideanj}
\end{equation}
In this case, the coalgebraic left and right quantum integrals \eqref{left1} arise as the sum of the squares of   components of the Dunkl angular momentum (\ref{momentaangdep}), namely:
\begin{equation}
	\hat{C}^{[m]} =\sum_{1 \leq i<j }^m  \,\hl_{ij}^2 \, , \quad 	\hat{C}_{[m]} =\sum_{N-m+1 \leq i<j }^N \,\hl_{ij}^2 \, . 
	\label{dunklang}
\end{equation}
This result is equivalent to the one obtained through coalgebra symmetry in quantum mechanics by using the common differential realization with standard momenta (without reflection operators). Here, as expected, the Dunkl angular momentum components take place of the standard angular momentum $\hat L_{ij}$ (\ref{standardL}), which are recovered when all $\mu_i$'s vanish. Notice that for $m=N$ we have that
\begin{equation}
	\hat{C}^{[N]} = \hat{C}_{[N]} = \sum_{1 \leq i<j }^N  \,\hl_{ij}^2  =:\hat{\boldsymbol{\Lambda}}^2 \, ,
	\label{eq:casn}
\end{equation}
where
\begin{equation}
	\hat{\boldsymbol{\Lambda}}^2 \Psi(\bx)=\hbar^2\left(-\bx^2 \Delta_\D+(\bx \cdot \nabla_\D)^2+(\bx \cdot \nabla_\D)\left((N-2)+2\sum_{i=1}^N \mu_i \hat{R}_i\right)\right)\Psi(\bx) \, ,
	\label{expcasn}
\end{equation}
represents the squared Dunkl angular momentum operator in dimension $N$.

The two specific choices of the central potential (\ref{HOKC})  make the system (\ref{eq:Euclideanj})   MS, leading to   the two celebrated Dunkl models that belong to this class, i.e.~the Dunkl oscillator and the Dunkl-KC system, which will be presented in Sections~\ref{subsec4.1} and \ref{subsec5.1}, respectively. If the $N$ non-central terms are added to the Dunkl oscillator 
and $N-1$ ones to the Dunkl-KC system, the generalized   Dunkl oscillator  and quasi-generalized Dunkl-KC system are thus obtained, which will be addressed, in this order, in Sections~\ref{subsec4.2} and \ref{subsec5.2}.
Therefore, these systems are MS cases of the Euclidean Dunkl family (\ref{eq:Euclideanjaddterms})  with quantum integrals quadratic in the momenta.


\subsection{Systems of  Dunkl type with central potentials on curved spaces} 
\label{sec31}
\noindent 
The  QMS Euclidean Dunkl family (\ref{eq:Euclideanjaddterms}) can be extended to curved spaces, with generically non-constant curvature,  through the so-called $\mathfrak{sl}(2,\mathbb{R})$-coalgebra spaces~\cite{sl2spaces,annals2009}.  We restrict ourselves to spherically symmetric spaces $\mani^N$ with metric given by
\be
\text{d} \metric^2=\ff(|\bx|)^2\text{d}\bx^2\, ,\quad |\bx|=\sqrt{\bx^2}\, ,\quad \text{d}\bx^2=\sum_{i=1}^N \dd x_i^2 \, ,
\label{zb}
\ee
where $\ff(|\bx|)$ is an arbitrary smooth function,  i.e.~the conformal factor of the Euclidean metric  $\text{d} \metric^2=\dd\bx^2$, and the variable $|\bx|$ is just a radial coordinate. The scalar curvature $\RR$ of the metric  (\ref{zb}) is determined by the conformal factor and the dimension of the manifold, turning out to be 
\be
\RR(|\bx|)=-(N-1)\,\frac{    (N-4)\ff '(|\bx|)^2+ \ff(|\bx|)  \left(    2\ff''(|\bx|)+2(N-1)|\bx|^{-1}\ff'(|\bx|)  \right)}   {\ff(|\bx|)^4  }\,   .
\label{zc}
\ee

As is well-known,  the family $\mani^N$ comprises the  three classical Riemannian spaces of   constant   curvature $\kappa$, denoted globally by $\mani^N_\kappa$, i.e.~the sphere $\bS^N$ $(\kappa>0)$,   hyperbolic   $\bH^N$ $(\kappa<0)$ and Euclidean $\bE^N$ $(\kappa=0)$ spaces. They naturally emerge by considering that the coordinates $\bx$ are Poincar\'e projective coordinates coming from  stereographic projection, which can be  constructed  in a unified form  for these spaces in terms of $\kappa$. For further applications in these spaces, it is worthy   explicitly recalling this construction~\cite{BH2007}. Let us consider the embedding of $\mani^N_\kappa$ in $\bR^{N+1}$ with ambient or Weierstrass coordinates $(\amb_0,\amb_1,\dots, \amb_N)=(\amb_0,\ambb)$ fulfilling the  ``sphere" constraint 
\be
\Sigma_\kappa:\ \amb_0^2+\kappa \ambb^2=1\, ,
\label{constraint}
\ee
and such that the origin   in $\mani^N_\kappa$ is taken as the ``north" pole $O=(1,\boldsymbol{0})\in \bR^{N+1}$. Hence, for $\kappa>0$ we get a proper sphere, but when $\kappa<0$ we find the two-sheeted hyperboloid. The flat contraction  $\kappa= 0$ gives rise to two Euclidean hyperplanes with $\amb_0=\pm 1$. Since $O=(1,\boldsymbol{0})$, hereafter it will be assumed that we are dealing with   the ``upper" sheet of the hyperboloid with $\amb_0\ge 1$ and the Euclidean hyperplane with $\amb_0= 1$. The metric on $\mani^N_\kappa$  reads as 
\be
\text{d}
\metric^2= \frac 1 \kappa
\left(\dd \amb_0^2+\kappa  \dd \ambb^2\right)\biggr|_{\Sigma_\kappa} \, .
\label{zd}
\ee
The stereographic projection with ``south" pole   $(-1,\boldsymbol{0})\in \bR^{N+1}$  leads to the  
Poincar\'e  projective coordinates $\bx$ from the   ambient coordinates $(\amb_0,\ambb)\in \Sigma_\kappa$, which is thus given by $(\amb_0,\ambb)=(-1,\boldsymbol{0})+\parp (1,\bx)\in \Sigma_\kappa$:
\be
\begin{split}
\parp&=\frac{2}{1+\kappa \bx^2}\, ,\quad \amb_0=\parp-1=\frac{1-\kappa\bx^2 }{1+\kappa\bx^2}\, ,\quad
\ambb= \parp\bx=\frac{2 \bx}{1+\kappa\bx^2}\, \\ \bx&=\frac{\ambb}{1+\amb_0} \, ,\quad\ \
\bx^2=\frac{1-\amb_0}{\kappa(1+\amb_0)} \, .
\end{split}
\label{ze}
\ee
Note that the  projection is well defined for any point in $\Sigma_\kappa$ (\ref{constraint}) except for the south pole  $(-1,\boldsymbol{0})$ which goes to points at infinity, $\bx\to\infty$, in both $\bS^N$  and $\bH^N$, meanwhile the origin  (north pole)   $O=(1,\boldsymbol{0})\in  \Sigma_\kappa$ is mapped on  the origin $ \boldsymbol{0}\in \mani^N_\kappa$. In addition, from (\ref{ze}) we find that:
\begin{itemize}
\item On $\bS^N$ with $\kappa>0$, it is verified that  $-1<\amb_0\le 1$ so that the domain of the Poincar\'e coordinates has no restrictions, $\bx\in(-\infty,+\infty)$, the projection maps the equator with $\amb_0=0$ on  the $N$-ball $\bx^2=1/\kappa$, the northern hemisphere with $\amb_0>0$ on the region inside the ball $\bx^2<1/\kappa$, and the southern hemisphere with $\amb_0<0$ on the outside region $\bx^2>1/\kappa$.

\item By contrast, on $\bH^N$ with $\kappa=-|\kappa|<0$ and $\amb_0\ge 1$,  the Poincar\'e coordinates are restricted to the inner region of the $N$-ball   with radius  $1/ |\kappa|$,
\be
\bx^2=\frac{ \amb_0-1}{|\kappa|( \amb_0+1)}< \frac{1}{|\kappa|}\, ,
\label{ballH}
\ee
which for $N=2$ it is just the interior of the  Poincar\'e disk (or conformal disk model), and  
whose boundary corresponds to the points at infinity $\amb_0\to \infty$. 
 
\item Finally, on the flat  $\bE^N$ with $\kappa=0$ and $\amb_0=+1$, the ambient coordinates $\ambb$ become   Cartesian coordinates with the metric (\ref{zd}) reducing to 
 $ \text{d} \metric^2=   \dd \ambb^2$.  Therefore, in this case, the Poincar\'e coordinates $\bx$ (\ref{ze}) are related to
the Cartesian ones by a factor ``$1/2$":  $ \bx = \ambb/ 2$. 
\end{itemize}
This discussion becomes quite relevant when solving a specific Dunkl type system   in $\bS^N$ or
 $\bH^N$ as strong differences arise between their spectrum and eigenfunctions (see, e.g., \cite{Kuru2016} for quantum oscillators for $N=2$ on these spaces).

 Therefore, under the parametrization (\ref{ze}),  the metric (\ref{zd}) becomes a particular case of (\ref{zb}) with conformal factor and constant scalar curvature (\ref{zc}) given by
\be
\text{d} \metric^2=\frac{4\dd \bx^2}{ (1+\kappa\bx^2 )^2} \, , \quad f(|\bx|) =\frac{2}{1+\kappa \bx^2} \, , \quad \RR(|\bx|)=N(N-1)\kappa \, ,
\label{zf}
\ee
comprising the three spaces $\mani^N_\kappa$.
 In $\bE^N$ with $\kappa=0$, the metric reduces to $\text{d} \metric^2=   \dd \ambb^2= {4\dd \bx^2}$ due to the factor ``1/2" in the Poincar\'e coordinates mentioned above. For this reason, and for the sake of simplicity, we will eliminate this factor when dealing with    Hamiltonian systems on $\bS^N$ and $\bH^N$, thus considering as conformal factor $f(|\bx|) = (1+\kappa|\bx|^2)^{-1}$ which yields $f(|\bx|) =1$ under the flat limit $\kappa=0$, recovering the initial Euclidean metric  $\text{d} \metric^2=\dd\bx^2$ in (\ref{zb}); in fact, this can be  regarded simply as a rescaling of the Hamiltonian.

In the classical framework, the metric (\ref{zb}) provides a family of free Hamiltonians describing the corresponding geodesic motion, which can be expressed  in terms of  the generators of the Poisson $\mathfrak{sl}(2,\mathbb{R}) $ coalgebra in the form~\cite{sl2spaces,annals2009}
\be
 {\TT} = \frac { {J}_+ }{ 2 \ff\big( \sqrt{{J}_- }\,\big)^2 }  = \frac{\bp^2}{2\ff (|\bx| )^2}\, .
 \label{zg}
\ee
However, ordering problems appear in the quantum case and there are different procedures to 
 construct the free quantum Hamiltonians. One approach makes use of the usual Laplace-Beltrami operator $\Delta_{\mathrm{LB}}$ determined by the metric (\ref{zb})  (see, e.g., \cite{Kalnins2003}). A second procedure consists in defining the quantization of the classical system (\ref{zg}) through the conformal (or Yamabe) Laplacian  $\Delta_{\mathrm{c}}$~\cite{Liu1992,Bar2003,Michel2014}, which is related to $\Delta_{\mathrm{LB}}$ via a term proportional to the scalar curvature as
 \be
 \Delta_{\mathrm{c}}= \Delta_{\mathrm{LB}}-\frac{N-2}{4(N-1)}\RR \, .
  \label{zh}
 \ee
 Therefore, the two quantizations coincide  for $N=2$ and differ by a constant when $\RR$ is constant as in the spaces $\mani^N_\kappa$  (\ref{zf}). Another procedure is the so-called ``direct Schr\"odinger quantization", which requires to place the conformal factor on one side of the Laplacian $\Delta$, namely
  \be
 \hat{\TT} = \frac { 1 }{ 2 \ff  \big(  {\hat{J}_-^{1/2} } \big)^2 }\, \hat{J}_+ = \frac{1}{2\ff (|\hbx| )^2}\, \hbp^2=
 -\frac{\hbar^2}{2\ff  (  |\bx|  )} \Delta
  \, .
  \label{zi}
  \ee
   In this paper we will consider the last prescription since it facilities the computations and, moreover, the quantum Hamiltonian $ \hat{\TT}$ can be related to the quantum conformal $ \hat{\TT}_{\text{c}}$, defined through (\ref{zh}), by means of a similarity transformation. For more details on these quantization procedures and their relationships we refer to~\cite{BEHRR2011,BEHRR2014}.
   
       Consequently, if we consider the quantum free Hamiltonian (\ref{zi}) in a Dunkl setting, introducing the differential-difference realization (\ref{NDrepquantum}),   and add a central potential, we obtain the curved counterpart of the flat Dunkl type Hamiltonian (\ref{eq:Euclideanjaddterms}) given by the following particular function $F\big(\hat J_+,\hat  J_-,\hat  J_3\big)$ (\ref{eq:ham}):
     \begin{align}
\hat{H} = \frac { 1 }{ 2 \ff  \big(  {\hat{J}_-^{1/2} } \big)^2 }\, \hat{J}_+   +  V\!\left(\!\sqrt{\hat{J}_-} \,\right)&= \frac { 1 }{ 2 \ff(|\hbx| )^2 }\left(  {\hbpi^2} + \sum_{i=1}^N \frac{\beta_i+\gamma_i \hr_i}{ \hx_i^2} \right)+ V(|\hbx|)\nonumber \\
&=\frac { 1 }{ 2 \ff(|\bx| )^2 }\left(  - {\hbar^2} \Delta_\D+   \sum_{i=1}^N \frac{\beta_i+\gamma_i \hr_i}{ x_i^2} \right)+V(|\bx|)
\, .
\label{eq:curvedaddterms}
\end{align}
This result can be seen as  an infinite family of curved Hamiltonians of Dunkl type with a central potential on spherically symmetric spaces $\mani^N$ (\ref{zb}),  which according to Proposition~\ref{prop1}  are all  QMS  sharing again the set of $2N-3$ universal quantum integrals (\ref{left1}). Clearly, specific choices of the conformal factor and the potential yield MS systems with additional integrals which, in general, are of higher-order in the momenta. We will focus on MS Hamiltonians with quantum symmetries quadratic in the momenta, identifying simultaneously the underlying curved space and the central potential. In particular, we construct the Dunkl oscillator and the Dunkl-KC system on the spaces of constant curvature $\mani^N_\kappa$ in Sections~\ref{subsec4.3} and \ref{subsec5.3}, together with their (quasi-)generalization (with curved centrifugal-reflection non-central potentials) in Sections~\ref{subsec4.4} and \ref{subsec5.4}, respectively. Furthermore, two MS systems which can be interpreted as a
Dunkl oscillator and a  Dunkl-KC system on spaces of non-constant curvature are addressed, in this order, in Sections~\ref{subsec4.5} and \ref{subsec5.5}, and, finally,  their (quasi-)generalization is obtained in Sections~\ref{subsec4.6} and \ref{subsec5.6}.


\section{Maximally superintegrable Dunkl oscillator-type systems}
\label{sec4}

\noindent In this section  we discuss some specific MS systems of Dunkl oscillator-type encompassed by the (flat) Euclidean and curved QMS families (\ref{eq:Euclideanjaddterms}) and (\ref{eq:curvedaddterms}). Note that, clearly, the latter leads 
to the Euclidean systems by setting $\ff(|\hbx| )=1$. The steps of our procedure can be summarized as follows:
\begin{itemize}
\item Consider (curved) oscillators which are already known to be MS with quadratic integrals in the usual quantum context and  which can be constructed by means of a function $F\big(\hat J_+,\hat  J_-,\hat  J_3\big)$ under the realization  (\ref{repr0}) (see Remark~\ref{rem4}). This, in turn, means that they share an $\mathfrak{sl}(2, \mathbb{R})$ coalgebra symmetry and, moreover, possess a (curved) Demkov-Fradkin type  tensor or, at least, some generalization of its diagonal components.

\item Extend them to the Dunkl setting by Proposition~\ref{prop1}, i.e., to a particular system contained in   (\ref{eq:curvedaddterms}), by keeping the same formal  function $F\big(\hat J_+,\hat  J_-,\hat  J_3\big)$  but now expressed under the realizations \eqref{NDrepquantumdunkl}, involving the parameters $\mu_i$, and (\ref{NDrepquantum}) with additional arbitrary parameters  $\beta_i$ and  $\gamma_i$.

\item And, finally,  explicitly deduce  the corresponding Dunkl-Demkov-Fradkin type tensor, or quantum integrals that extend its diagonal components,  which for curved systems is by no means a trivial task, thus proving the MS property.

 \end{itemize}

\noindent In this way,  we will observe that most of the Dunkl  oscillator-type models appearing in the literature share the same   $\mathfrak{sl}(2, \mathbb{R})$ coalgebra symmetry, and we will retrieve them as particular MS cases. In addition, 
we  will go further by presenting some new models which, to the best of our knowledge, are not present in the literature yet.


\subsection{Dunkl oscillator}
\label{subsec4.1}

\noindent 
The proper  Dunkl oscillator system arises with the following choice of the function (\ref{eq:ham}):
\begin{equation}
F\big(\hat J_+,\hat  J_-,\hat  J_3\big)= \frac{\hat  J_+}{2} + \omega^2 \frac{\hat  J_-}{2} \,  \quad (\omega \in \mathbb{R}) 
\label{eq:dunkl-osc}
\end{equation}
through the $N$D realization of  $\mathfrak{sl}(2, \mathbb{R})$ \eqref{NDrepquantumdunkl}. This corresponds to taking $V \big(\hat  J_-\big)= \omega^2  \hat  J_-/2$ in \eqref{eq:Euclideanj}, i.e., $V_{\textsf{HO}}$ (\ref{HOKC}).
The $N$D Hamiltonian, obtained under the above realization, is thus given by 
\begin{equation}
\hat{H} = \frac{\hbpi^2}{2}+ \frac{\omega^2\hbx^2}{2} = -\frac{\hbar^2}{2}\Delta_\D+\frac{\omega^2\bx^2}{2} \, .
\label{eq:dunkl}
\end{equation}
 Since the system is coalgebraic, the left and right quantum
integrals \eqref{dunklang} automatically commute with the Hamiltonian. This $N$D quantum Hamiltonian system turns out to be MS. In particular, there exists a Dunkl version of the Demkov-Fradkin tensor \cite{D1959,F1965}  involving reflections. Its components:
\begin{equation}
\hat{\mathcal{F}}_{ij} =  \hpi_i \hpi_j +\omega^2 \hx_i \hx_j  \, , \quad i, j=1, \dots, N
\label{eq:demkovfradkin}
\end{equation}
individually commute with the Hamiltonian operator \eqref{eq:dunkl}, 
\begin{equation}
\big[\hat{H}, 	\hat{\mathcal{F}}_{ij} \big]=0 \, , \quad i,j=1, \dots, N\, ,
\label{eq:hamhop}
\end{equation}
which can be  easily checked. Let us underline that these additional quantum integrals are not directly provided by the coalgebra symmetry of the model, but have to be found by recurring to other ways.

  Notice that, as for the standard $N$D quantum harmonic oscillator, the Hamiltonian of the $N$D Dunkl oscillator can be rewritten as 
 \begin{equation}
 \hat{H} =\frac 12  \sum_{i=1}^N\hat{\mathcal{F}}_{ii}\, ,
 \label{traceDF}
 \end{equation}
 showing  separability in Cartesian coordinates.
	 
\begin{rem}
\label{rem41}
For $N= 2$, we recover the model studied in \cite{GIVZ2013I,GIVZ2013II},  whereas for $N = 3$ the one presented in \cite{GVZ2014}. The $N$D case can be found in \cite{Ghaz2021}.
\end{rem}


\subsection{Generalized Dunkl oscillator: Dunkl-Smorodinsky-Winternitz system}
\label{subsec4.2}

\noindent This model emerges while preserving the same formal function (\ref{eq:dunkl-osc}), but now expressed under the (complete) $N$D realization \eqref{NDrepquantum} giving rise to the following  quantum Hamiltonian operator 
\begin{equation}
\hat{H} = \frac{\hbpi^2}{2}+\sum_{i=1}^N \frac{\beta_i +\gamma_i \hr_i}{2\hx_i^2}+ \frac{\omega^2\hbx^2}{2} = -\frac{\hbar^2}{2}\Delta_\D+\sum_{i=1}^N \frac{\beta_i +\gamma_i \hr_i}{2x_i^2}+\frac{\omega^2\bx^2}{2} \, .
\label{singdunkl}
\end{equation} 
Also this system, due to its separability in Cartesian coordinates, turns out to be MS. In particular, besides the $2N - 3$ quantum integrals \eqref{left1} coming from coalgebra symmetry, we have another set of $N$ additional quantum integrals:

\begin{equation}
\hat{\mathcal{F}}_{i}=  {\hpi_i^2} +\frac{\beta_i +\gamma_i \hr_i}{ \hx_i^2}+  {\omega^2\hx_i^2}  = - {\hbar^2} D_i^2+\frac{\beta_i +\gamma_i \hr_i}{x_i^2}+ {\omega^2x_i^2} \, ,\, \quad i=1, \dots, N\, ,
\label{eq:sepconst}
\end{equation}
which determine 1D Hamiltonians and verify that
\begin{equation}
\hat{H}=\frac 12 \sum_{i=1}^N\hat{\mathcal{F}}_{i}\, ,\quad 
\big[\hat{H},\hat{\mathcal{F}}_{i}\big] =0 \, , \quad  i=1, \dots, N  \, .
\label{eq:commhi}
\end{equation} 
Hence, when $\{  \beta_i,  \gamma_i\}_{i=1}^N =0$, these integrals reduce to the diagonal components of the   Dunkl-Demkov-Fradkin tensor (\ref{eq:demkovfradkin}).

In the standard quantum case, with $\{  \mu_i,  \gamma_i\}_{i=1}^N =0$, we recover the isotropic oscillator with $N$ centrifugal $\beta_i$-terms, which is known as the  Smorodinsky-Winternitz system~\cite{WSUF65,Evans90,Evans91,GPS1995}. For this reason, the quantum Hamiltonian (\ref{singdunkl}) can be called ``Dunkl-Smorodinsky-Winternitz system". We recall that, in the 
standard classical setting, each $\beta_i$-term corresponds to the presence of a centrifugal (infinite) barrier which limits the trajectory of the particle on $\bE^N$. For instance, for $N=2$, if $\beta_1\ne 0$ and $\beta_2=0$ the trajectory is constrained to one half of the Euclidean plane $(x_1,x_2)$, whereas
for both $\beta_1$, $\beta_2$ different from zero, it is restricted to a quadrant.
In this sense, the  Smorodinsky-Winternitz system represents a particular case of the ``caged" oscillator~\cite{VerrierOscillator}. However, the behaviour  of the $\beta_i$'s and $\gamma_i$'s non-central potentials is somewhat   different as the latter are associated with  reflections but not with constants.

\begin{rem}
\label{rem42}
For $N= 2$, we recover the model appearing in \cite{GVZ2013}, where it was referred to as the singular oscillator. Notice that, in that paper, the parameters $\alpha_{x}, \alpha_{y}$ and $\beta_{x}, \beta_{y}$  (which correspond to our $\beta_1, \beta_2$ and $\gamma_1, \gamma_2$, respectively) satisfy appropriate quantization conditions arising from the parity requirements, due to the presence of reflections, on the solutions of the Schr{\"o}dinger equation associated with the Hamiltonian. This is to emphasize that, when the Schr{\"o}dinger equation is considered, additional constraints for the parameters have to be taken into account.
\end{rem}

\begin{rem}
\label{rem42b}
We observe that the proper Dunkl oscillator system (\ref{eq:dunkl}) can be rewritten by means of the gauge transformation introduced in~\cite{GLVZ2013} in the form
\begin{equation}
\hat{H} =  -\frac{\hbar^2}{2}\Delta +\frac{\omega^2\bx^2}{2} +\frac{1}{2}\sum_{ i=1}^N \frac {\mu_i^2}{x_i^2}-\frac{1}{2}\sum_{ i=1}^N \frac {\mu_i \hr_i}{x_i^2}\, .
\label{eq:dunklA}
\end{equation} 
Clearly, and as mentioned  in~\cite{GLVZ2013}, the  symmetry algebra  and the spectrum of the  Dunkl oscillator Hamiltonian are left. Thus,  the Dunkl-Smorodinsky-Winternitz Hamiltonian (\ref{singdunkl}) determines a  different system which should be solved along the lines performed in  \cite{GVZ2013}   commented above.
\end{rem}


\subsection{Dunkl oscillator on the sphere and hyperbolic space: Dunkl-Higgs system}
\label{subsec4.3}

 \noindent 
 We construct the Dunkl oscillator  on the three spaces $\mani^N_\kappa$ (\ref{zf}) in terms of the curvature $\kappa$, comprising the sphere $\bS^N$, hyperbolic $\bH^N$ and Euclidean $\bE^N$ spaces in a unified form. Hence, as a byproduct, the proper Dunkl oscillator in Section~\ref{subsec4.1} will be recovered through the flat contraction $\kappa=0$.
 
 With this aim, let us consider the Poincar\'e projective coordinates (\ref{ze}),  the (scaled) conformal factor 
 $f(|\bx|) = (1+\kappa|\bx|^2)^{-1}$ and the direct Schr\"odinger quantization (\ref{zi}) for the free quantum Hamiltonian obtaining
 \be
\hat \TT^{(\kappa)}=\big(1+\kappa \hat J_- \big)^2\frac{\hat J_+}{2} \, .
 \label{TTk}
 \ee
   The isotropic oscillator on $\mani^N_\kappa$ expressed in Poincar\'e coordinates turns out be~\cite{BH2007,annals2009}
\be
V^{(\kappa)}_{\textsf{HO}} = \frac{ \omega^2}{2} \frac{\hat{J}_-}{ \big( 1-\kappa \hat{J}_- \big)^2} =\frac  {\omega^2}{2}\frac{\hbx^2}{\big( 1-\kappa \hbx^2 \big)^2 } \, \quad (\omega, \kappa \in \mathbb{R}) \, ,\\
 \label{z43a}
\ee
reducing to the Euclidean $V_{\textsf{HO}}$ (\ref{HOKC}) for $\kappa=0$.  Recall that this expression comes, in fact, 
  from $V^{(\kappa)}_{\textsf{HO}}$ written in the $N+1$ ambient coordinates $(\amb_0,\ambb)$, subjected to the constraint (\ref{constraint}), by introducing its  parametrization  (\ref{ze}) in terms of the $N$ ``intrinsic" Poincar\'e  coordinates $\bx$, namely   
\be
V^{(\kappa)}_{\textsf{HO}} = \frac{ \tilde \omega^2}{2}\frac{\ambb^2}{\amb_0^2}= \frac{\tilde \omega^2}{2}\frac{1-\amb_0^2}{\kappa\amb_0^2}=\frac  {\omega^2}{2}\frac{\bx^2}{ ( 1-\kappa \bx^2  )^2 }\, , \quad \tilde \omega  = \frac{\omega }2\, .
\label{kHO}
\ee  
Therefore, we have a spherical/hyperbolic oscillator potential, which can be expressed  in a more  common way in terms of the  distance $r$ along the geodesic $\ell$ joining the particle and the origin in $\mani^N_\kappa$, which does not coincide with the radial coordinate $|\bx|$ used in hyperspherical coordinates;  the geodesic radial coordinate $r$ actually corresponds to
 \be
    \amb_0=\cos\big(\sqrt{\kappa}\,r\big)\,  , \quad |\bx|=\frac{1}{\sqrt{\kappa}}\tan\!\left(\sqrt{\kappa}\,\frac {r}{2}\right) \, .
    \label{geod}
\ee 
      Explicitly, for normalized values $\kappa=\pm 1$, the resulting potentials and   transformations of the   radial variable $ r$ are given by 
\be
\begin{aligned}
\bS^N \ (\kappa=+1)\!:&  \quad V^{(+)}_{\textsf{HO}} =\frac 12\,\tilde \omega^2  \tan^2\! r\,  ,\quad\  \,   \amb_0= \cos r  \, ,\quad\ \      |\bx|= \tan \frac { r}2\, \\[2pt]
\bH^N \ (\kappa=-1)\!:&\quad  V^{(-)}_{\textsf{HO}} =\frac 12\,\tilde \omega^2  \tanh^2\! r\,  ,\quad  \amb_0= \cosh r  \, ,\quad  |\bx|= \tanh \frac { r}2\, ,
\end{aligned}
\label{HiggsSS}
\ee
  where  $V^{(+)}_{\textsf{HO}} $ is just the (spherical) Higgs  oscillator  \cite{Higgs1979, Leemon1979}  in its usual form and we can then call $V^{(-)}_{\textsf{HO}}$ the hyperbolic Higgs oscillator  (see, e.g., \cite{Car2004,Santander6} and references therein).

  In this way, the complete
 ``Dunkl-Higgs oscillator" on $\mani^N_\kappa$ follows from (\ref{TTk}) and (\ref{z43a}), i.e., 
by  the function
\begin{equation}
F\big(\hat J_+,\hat J_-,\hat J_3\big)= \big(1+\kappa \hat J_- \big)^2\frac{\hat J_+}{2} +\frac{\omega^2}{2}\frac{\hat  J_-}{\big(1-\kappa \hat J_-\big)^2}\,    \quad (\omega, \kappa \in \mathbb{R}) \, ,
\label{eq:quantumHiggs}
\end{equation}
which, under  the $N$D realization of the $\mathfrak{sl}(2, \mathbb{R})$ \eqref{NDrepquantumdunkl}, yields the Hamiltonian
 \begin{equation}
\hat{H}^{(\kappa)} =\big(1+\kappa \hbx^2 \big)^2\frac{\hbpi^2}{2}+\frac{\omega^2}{2}\frac{\hbx^2}{\left(1-\kappa \hbx^2\right)^2} = -\frac{\hbar^2(1+\kappa \bx^2)^2}{2}\Delta_\D+\frac{\omega^2}{2}\frac{\bx^2}{(1-\kappa \bx^2)^2} \, .
\label{eq:DHiggs}
\end{equation}
The corresponding left and right quantum integrals are given by \eqref{dunklang}. To show that $\hat{H}^{(\kappa)}$ is MS, it is necessary to obtain at least one  additional integral  that gives  a set of  $2N-1$ algebraically independent operators, including $\hat{H}^{(\kappa)}$ and the  $2N-3$  \eqref{dunklang}.
 For oscillators, this means deriving some kind of Demkov-Fradkin tensor (or some generalization of its diagonal components).
 Let us define  appropriate curved spherical/hyperbolic  Dunkl  momentum operators as
 \begin{equation}
\hat{\Gamma}_i^{(\kappa)}:=\big(1-\kappa \hbx^2\big)\hpi_i+2 \kappa  \hat{x}_i \left((\hbx \cdot \hbpi)-\imath \hbar \sum_{  j=1}^N \mu_j \hat{R}_j\right) \, , \,  \quad i=1, \dots, N .
\label{eq:operator}
\end{equation}
 It should be noted that these operators behave as genuine  
Dunkl  momenta on $\bS^N$ and $\bH^N$ according to the commutators between them and with the Dunkl  angular momentum operators $\hl_{ij} $. The corresponding  algebraic structure,  including also the reflection operators $\hr_{i}$, is discussed in detail in Appendix~\ref{AA}, giving rise to a quadratic algebra which enlarges the algebra $\mathfrak{so}\big(N, \mu_1 \hr_1, \dots, \mu_N  \hr_N\big)$   introduced in  \cite{Ghaz2021}  (see  Remark~\ref{rem1}). 

 From $\hat{\Gamma}_i^{(\kappa)}$, we construct the  following $N$  operators 
  \begin{equation}
\hat{\mathcal{I}}_i^{(\kappa)}=\frac{1}{2}\left(\hat{\Gamma}_i^{(\kappa)}\right)^2+\frac{\omega^2}{2}\frac{\hat{x}_i^2}{\left(1-\kappa \hbx^2\right)^2}+(N-2)\hbar \kappa \left( -\imath \hat{x}_i \hat{\Gamma}_i^{(\kappa)}+\hbar \kappa \left(\hbx^2 \mu_i \hr_i +\frac{1}{2}\left( \hbx^2-N\hat{x}_i^2 \right)\right) \right)\, ,
\label{constcurvedhiggs}
\end{equation}
which turn out to be quantum integrals of $\hat{H}^{(\kappa)}$  (\ref{eq:DHiggs}),  i.e., it can be verified that
\begin{equation}
\big[\hat{H}^{(\kappa)}, \hat{\mathcal{I}}_{i}^{(\kappa)}  \big]=0\, ,\quad  i=1, \dots, N\, .
 \label{eq:commrelI}
\end{equation}
The MS property of  $\hat{H}^{(\kappa)}$ is established  by choosing  an operator $\hat{\mathcal{I}}_i^{(\kappa)}$ (fixing the index $i$),  the    $2N-3$    integrals \eqref{dunklang} together with  $\hat{H}^{(\kappa)}$, since the set  $\big\{\hat{H}^{(\kappa)}, \hat{C}^{[m]},\hat{C}_{[m]},\hat{\mathcal{I}}_i^{(\kappa)}\big\}$, with $m=2,\dots, N$, consists of 
$2N-1$ algebraically independent operators; recall the relation (\ref{eq:casn}) and observe that $\hat{\mathcal{I}}_i^{(\kappa)}$ only contains a single Dunkl momentum $\hpi_i$.

Consequently,  this system represents a MS extension by reflections of the Higgs oscillator on both $\bS^N$ and $\bH^N$.  Let us   notice that, under the flat $\kappa \to 0$ limit,   the  curved Dunkl  momentum operators $\hat{\Gamma}_i^{(\kappa)}$ (\ref{eq:operator}) reduce to the usual Euclidean Dunkl ones, namely $\hat{\Gamma}_i^{(0)}=\hpi_i$, and the quantum integrals \eqref{constcurvedhiggs} collapse to the diagonal components of the Dunkl-Demkov-Fradkin tensor \eqref{eq:demkovfradkin}.  In addition,  the Hamiltonian \eqref{eq:DHiggs} can also be  expressed as:
\begin{equation}
\hat{H}^{(\kappa)}=\sum_{ i=1 }^N  \hat{\mathcal{I}}_i^{(\kappa)}+2\kappa \left(\hat{\boldsymbol{\Lambda}}^2+\hbar^2 \left(\sum_{i=1}^N \mu_i\left(\mu_i+\frac{N}{2}\hat{R}_i\right)+2\sum_{1 \leq i <j}^N \mu_i \mu_j \hat{R}_i \hat{R}_j \right)\right) \, ,
\label{HamHiggs}
\end{equation}
where the operator $\hat{\boldsymbol{\Lambda}}^2$ is given in \eqref{eq:casn} and  \eqref{expcasn}. This expression clearly highlights the strong effect of the curvature $\kappa$, which is worth  comparing with (\ref{traceDF}).

\begin{rem}
\label{rem43}
We recall that Dunkl models on spaces with costant curvature have already been considered, in low dimensions, in the works \cite{NajaPana, DNPCH}   using geodesic polar coordinates~\cite{BHSZG,HBSSG}. 
  In this respect, we note that in order to extend the proper  Dunkl oscillator  to the sphere and hyperbolic spaces in arbitrary dimension, Poincar\'e projective variables  arise naturally within an $\mathfrak{sl}(2,\mathbb{R})$ coalgebra symmetry and facilitate  the  derivation of  explicit quantum integrals.
\end{rem}


\subsection{Generalized Dunkl oscillator on the sphere and hyperbolic space}
\label{subsec4.4}
\noindent 
Centrifugal-reflection non-central potentials can be added directly  to the Hamiltonian (\ref{eq:DHiggs}) keeping the same formal function (\ref{eq:quantumHiggs}), but now introducing the
$N$D realization (\ref{NDrepquantum}) with generic parameters $\{\beta_i , \gamma_i \}_{i=1}^N\ne 0$.   The Hamiltonian results in:
\begin{align}
\hat{H}^{(\kappa)} &=\frac{\big(1+\kappa \hbx^2 \big)^2}{2}	\left(\hbpi^2+\sum_{i=1}^N \frac{\beta_i+\gamma_i \hat{R}_i}{\hat{x}_i^2}\right)+\frac{\omega^2}{2}\frac{\hbx^2}{\left(1-\kappa \hbx^2\right)^2} \nonumber \\
&= \frac{(1+\kappa \bx^2)^2}{2}\left(-\hbar^2\Delta_\D+\sum_{i=1}^N \frac{\beta_i+\gamma_i \hat{R}_i}{x_i^2}\right)+\frac{\omega^2}{2}\frac{\bx^2}{(1-\kappa \bx^2)^2} \, .
\label{eq:curvedSW}
\end{align}
 The left and right quantum integrals are    given in this  case by the complete set \eqref{left1}. This quantum model becomes the generalization of the previous spherical/hyperbolic Dunkl oscillator through non-central potentials, which turns out to be again MS due to the existence of the following $N$ additional quantum integrals:
\begin{align}
\hat{\mathcal{J}}_i^{(\kappa)}&=\frac{1}{2}\left(\hat{\Gamma}_i^{(\kappa)}\right)^2+\frac{\big(1-\kappa \hbx^2 \big)^2}{2\hat{x}_i^2} \big(\beta_i+\gamma_i \hat{R}_i \big)+\frac{\omega^2}{2}\frac{\hat{x}_i^2}{\left(1-\kappa \hbx^2\right)^2} \nonumber\\
&\qquad +(N-2)\hbar \kappa \left( -\imath \hat{x}_i \hat{\Gamma}_i^{(\kappa)}+\hbar \kappa \left(\hbx^2 \mu_i \hr_i +\frac{1}{2}\left( \hbx^2-N\hat{x}_i^2 \right)\right) \right)\, ,
\end{align}
for $i=1, \dots, N$ and with the curved Dunkl momentum  $\hat{\Gamma}_i^{(\kappa)}$ defined in  (\ref{eq:operator}).  It can be checked, by means of cumbersome but simple calculations, that the following commutation relations are fulfilled: 
\begin{equation}
\big[\hat{H}^{(\kappa)}, \hat{\mathcal{J}}_{i}^{(\kappa)}  \big]=0\, ,\quad  i=1, \dots, N\, .
\label{eq:commrelJ}
\end{equation}
Similarly to the relation (\ref{HamHiggs}), the Hamiltonian \eqref{eq:curvedSW} can be expressed as:
\begin{equation}
	\hat{H}^{(\kappa)}=\sum_{i=1}^N  \hat{\mathcal{J}}_i^{(\kappa)}+2\kappa \left(\hat{C}^{[N]}+\hbar^2 \left(\sum_{i=1}^N \mu_i\left(\mu_i+\frac{N}{2}\hat{R}_i\right)+2\sum_{1 \leq i <j}^N \mu_i \mu_j \hat{R}_i \hat{R}_j \right)\right) \, ,
	\label{HamHiggs2}
\end{equation}
where $\hat{C}^{[N]}= \hat{C}_{[N]}$ is now given by \eqref{CN}.  Then, for $i$ fixed and $m=2,\dots,N$, the set  $\big\{\hat{H}^{(\kappa)}, \hat{C}^{[m]},\hat{C}_{[m]}, \hat{\mathcal{J}}_i^{(\kappa)}\big\}$ is formed by
$2N-1$ algebraically independent operators and, therefore, $\hat{H}^{(\kappa)}$ \eqref{eq:curvedSW} is  MS. To the best of our knowledge, the Hamiltonian  \eqref{eq:curvedSW}  represents a new oscillator model of Dunkl type.

Recall that the standard quantum Hamiltonian without reflections, obtained by setting $\{\mu_i , \gamma_i \}_{i=1}^N=0$ in \eqref{eq:curvedSW},  thus corresponding to the spherical/hyperbolic Higgs oscillator with the $N$ ``curved" centrifugal $\beta_i$-potentials, is also known as the ``curved Smorodinsky-Winternitz system"  \cite{BHSZG, HBSSG, Pogosyan1995}.  In this respect, we stress that such $\beta_i$-terms can alternatively be interpreted as non-central oscillators only on the $N$-sphere $\bS^N$, while they behave as usual (curved) centrifugal barriers on $\bH^N$~\cite{BHSZG, HBSSG,ran2003}. As $\hat{H}^{(\kappa)}$ (\ref{eq:curvedSW})  can be regarded as  an extension by reflections of the curved Smorodinsky-Winternitz system, it is worth describing in detail the role played by the $\beta_i$- and $\gamma_i$-potentials for such  ``curved Dunkl-Smorodinsky-Winternitz system".

 The Hamiltonian $\hat{H}^{(\kappa)}$  (\ref{eq:curvedSW})  can be split into the curved kinetic operator $\hat{\TT}^{(\kappa)}$ (\ref{TTk}) on the spaces $\mani^N_\kappa$   and a generalized curved Dunkl oscillator potential $\hat{V}_{\textsf{gHO}}^{(\kappa)}$ comprising the $\beta_i$- and $\gamma_i$-terms, which are eventually provided by the realization of $\hat J_+$, as
 \be
 \hat{H}^{(\kappa)}= \hat{\TT}^{(\kappa)}+\hat{V}_{\textsf{gHO}}^{(\kappa)}\, ,\quad
  \hat{\TT}^{(\kappa)} = -\frac{\hbar^2(1+\kappa \bx^2)^2}{2}\Delta_\D \,  , 
   \quad \hat{V}_{\textsf{gHO}}^{(\kappa)}= \hat{V}_{\textsf{HO}}^{(\kappa)}+ \frac{(1+\kappa \bx^2)^2}{2} \sum_{i=1}^N \frac{\beta_i+\gamma_i \hat{R}_i}{x_i^2} \, ,
   \label{gHOk}
 \ee
where $\hat{V}_{\textsf{HO}}^{(\kappa)}$ is the Higgs oscillator given in (\ref{z43a}). 
The potential $\hat{V}_{\textsf{gHO}}^{(\kappa)}$ can be written more  simply in terms of the $N+1$ ambient coordinates $(\amb_0,\ambb)$, subjected to the constraint  (\ref{constraint}). From the projective relations   (\ref{ze}) we find that
\be
\hat{V}_{\textsf{gHO}}^{(\kappa)}=  \frac{ \tilde \omega^2}{2}\frac{\ambb^2}{\amb_0^2}  + 2 \sum_{i=1}^N \frac{\beta_i+\gamma_i \hat{R}_i}{\amb_i^2} \, ,\quad \tilde \omega = \frac{\omega}{2}\, ,
   \label{gHOk2}
\ee
allowing the potential to be expressed in any set of $N$ intrinsic coordinates on $\mani^N_\kappa$ \cite{BHSZG, HBSSG}. Note that the notation is consistent since $[\hr_i,\hx^2_i]=[\hr_i,\hat \amb^2_i]= 0$. In the classical picture, let us consider the particle  located at a generic point  $P=(\amb_0,\ambb)\in \Sigma_\kappa$, from a radial  distance $r$ to the origin $O=(1,\mathbf{0})$ measured along the geodesic $\ell$ joining both points.  Let $\{\ell_i\}_{i=1}^N$ be a geodesic reference frame such that all these basic geodesics are mutually orthogonal  at $O$ and each $\ell_i$ is given by the intersection of $ \Sigma_\kappa$  with the coordinate plane determined by the axes $\{\amb_0,\amb_i\}$. Let $P_i$ be the point at a geodesic distance $\yy_i$ from the particle $P$, measured along the geodesic joining $P$ and $P_i$, and orthogonal to the set $\{\ell_j\}_{1=j\ne i}^N$ $(i=1,\dots, N)$. Then, by applying trigonometry on  $\mani^N_\kappa$ (see \cite{2013Higgs} for the explicit construction for $N=2$),   the ambient coordinates $(\amb_0,\ambb)$  can be written in terms of the geodesic distances $(r,\byy)$ as
\be
   \amb_0=\cos\big(\sqrt{\kappa}r\big)\,  , \quad   \amb_i=\frac1{\sqrt{\kappa}}\sin\big(\sqrt{\kappa}\,\yy_i\big)\,  ,  \quad 
   \label{geod2}
\ee
the former relation already given in (\ref{geod}). Thus,  for normalized values $\kappa=\pm 1$,  the generalized potential (\ref{gHOk2}) becomes
\be
\begin{aligned}
\bS^N \ (\kappa=+1)\!:&  \quad V^{(+)}_{\textsf{gHO}} =\frac 12\,\tilde \omega^2  \tan^2\! r +2\sum_{i=1}^N\frac{\beta_i+\gamma_i \hat{R}_i}{\sin^2\!\yy_i}  
   \\[2pt]
\bH^N \ (\kappa=-1)\!:&\quad  V^{(-)}_{\textsf{gHO}} =\frac 12\,\tilde \omega^2  \tanh^2\! r+2\sum_{i=1}^N\frac{\beta_i+\gamma_i \hat{R}_i}{\sinh^2\!\yy_i}   \, ,
\end{aligned}
\label{gHiggsSS}
\ee
showing the role of the $\beta_i$ and $\gamma_i$-terms as ``curved central barriers", which are proportional  to
$\yy_i^{-2}$ in the Euclidean case. In addition, only for the $N$-sphere  (take again $\kappa=+1$ for simplicity), there exist $N$ fixed points $O_i$ $(i=1,\dots, N)$ obtained from the intersection of each coordinate axis $s_i$ with the sphere $\amb_0^2+\ambb^2=1$. Therefore, all of them are located at the equator $\amb_0=0$, with ambient coordinates $\{O_1=(0,1,0,\dots,0),\dots,  O_N=(0,\dots,0,1)\}$,  
    mutually at a distance $\pi /2$,   and each $O_i$ also at a distance $\pi /2$ from the origin $O=(1,\mathbf{0})$ along the basic geodesic $\ell_i$. This fact enables us to introduce the spherical distance $r_i$ from the fixed point $O_i$ to the origin $O$, complementary to  $\yy_i$, such that $r_i+\yy_i=\pi/2$ and $\sin\yy_i=\cos r_i$. Consequently, choosing positive parameters $\beta_i=\omega_i^2/4>0$, 
     the generalized spherical Dunkl potential (\ref{gHiggsSS}) can be rewritten in the form
 \be
   V^{(+)}_{\textsf{gHO}} =\frac 12\,\tilde \omega^2  \tan^2\! r +\frac 12 \sum_{i=1}^N   \omega_i^2  \tan^2\! r_i  +2\sum_{i=1}^N \gamma_i\big(1+\tan^2\! r_i \big){  \hat{R}_i} +\frac{1}{2} \sum_{i=1}^N \omega_i^2 \, ,
   \label{gHOSN}
 \ee
  which can finally be interpreted as the superposition of the Higgs oscillator with centre at the origin $O$ in $\bS^N$ with $N$ non-central spherical oscillators with centres at the points $O_i$, together with $N$ spherical non-central oscillator-reflection potentials. This construction cannot be performed for $\bH^N$ as there are no points like  $O_i$   (these would be beyond the infinity in the ``ideal" region of the hyperbolic space).  This geometrical interpretation can be expected to have profound consequences and differences when solving the    Schr\"odinger  equation for the generalized spherical/hyperbolic Dunkl oscillators, which would be the curved counterpart of the results already obtained in \cite{GVZ2013}  for $\bE^2$ as pointed out in Remark~\ref{rem42}.   
  
\begin{rem}  
\label{rem44}
Observe that the  Hamiltonian  (\ref{eq:curvedSW}) does  not coincide with the so-called ``Smorodinsky-Winternitz system"  in \cite{DNPCH}. Indeed, in that work, the terminology comes from the $\mu_i$-terms appearing in the Dunkl momentum operators (\ref{dunklmom}), which are there interpreted as Rosochatius-Winternitz potentials. More precisely, these arise through a gauge transformation~\cite{GLVZ2013}, similarly to the Euclidean case (\ref{eq:dunklA})  in Remark~\ref{rem42b}. Hence, the Smorodinsky-Winternitz model in \cite{DNPCH} is nothing more than the Dunkl-Higgs oscillator for $N=3$ written in geodesic polar variables  as mentioned in Remark~\ref{rem43}.  In this respect, note also that if we drop all the non-central terms, setting $\{\beta_i , \gamma_i \}_{i=1}^N=0$, we recover the results obtained in Section \ref{subsec4.3}. If we instead keep them, and consider the flat $\kappa \to 0$ limit, we recover the results reported in Section \ref{subsec4.2}.
\end{rem}


\subsection{Dunkl-Darboux III system} 
\label{subsec4.5}

\noindent
Among the spaces $\mani^N$  (\ref{zb}) with a non-constant curvature,  let us now consider the so-called Darboux III space,  since it underlies a paradigmatic oscillator. This model is MS with quantum symmetries quadratic in the momenta, i.e., with a quantum Demkov-Fradkin tensor \cite{BEHRR2011,BEHRR2011b} (see \cite{Latini2016} for a complete description and solution of the classical system). The Darboux III space $\bD^N$ corresponds to choose the conformal factor
$\ff(|\bx|)=\sqrt{1+\lambda \bx^2}$, thus giving rise to the  following metric (\ref{zb}) and scalar curvature  (\ref{zc}) 
\be
\text{d} \metric^2=  (1+\lambda \bx^2 )\text{d} \bx^2\, , \quad \RR(|\bx|)= - \lambda (N-1)\,\frac{2N + 3\lambda(N-2)\bx^2}{(1+\lambda \bx^2)^3}\, ,
\label{spaceD}
\ee
where the parameter $\lambda\in \bR$, and can be seen as a ``deformation" parameter since when $\lambda=0$ the flat space $\bE^N$ is recovered. Notice that the sign of $\lambda$ is relevant. For $\lambda>0$, there is no restriction for the coordinates, $\bx\in (-\infty,+\infty)$, while if $\lambda=-|\lambda|<0$, the domain of $\bx$ is restricted to the interior of the $N$-ball $\bx^2 <1/ |\lambda|$, reminding the hyperbolic space $\bH^N$ with respect to the constant curvature $\kappa$,  as given in (\ref{ballH}).
This fact has deep consequences when solving any classical/quantum system on $\bD^N$, which can indeed be  interpreted as a spherical/hyperbolic space of non-constant curvature for  positive/negative values of $\lambda$.

Then, we apply the direct Schr\"odinger quantization for the quantum free  Hamiltonian (\ref{zi}) and add the Darboux III oscillator potential  with the following choice of the function $F$  (\ref{eq:ham}):
\begin{equation}
F\big(\hat J_+,\hat  J_-,\hat  J_3\big)= \frac{1}{1+\lambda \hat  J_-}\frac{\hat  J_+}{2} +\frac{\omega^2}{2}\frac{\hat  J_-}{1+\lambda\hat  J_-}  \quad (\omega, \lambda \in \mathbb{R}) \, , \quad 
\label{eq:quantumDIII}
\end{equation}
which, by introducing  the realization  \eqref{NDrepquantumdunkl}, yields the Dunkl-Darboux III Hamiltonian
\begin{equation}
	\hat{H}^{(\lambda)} = \frac{1}{1+\lambda \hbx^2}\frac{\hbpi^2}{2}+\frac{\omega^2}{2}\frac{\hbx^2}{1+\lambda \hbx^2} = -\frac{\hbar^2}{2(1+\lambda \bx^2)}\Delta_\D+\frac{\omega^2}{2}\frac{\bx^2}{1+\lambda \bx^2} \, .
	\label{eq:qghD}
\end{equation}
 The $2N-3$ left and right quantum integrals are given by \eqref{dunklang}. Furthermore, this $N$D quantum  oscillator model of Dunkl-type turns out to be MS thanks to the existence of a  curved version of the Dunkl-Demkov-Fradkin tensor \eqref{eq:demkovfradkin}, namely:
\begin{equation}
	\hat{\mathcal{F}}_{ij}^{(\lambda)} = \hpi_i \hpi_j + \hx_i \hx_j\left(\omega^2 -2 \lambda \hat{H}^{(\lambda)}\right)\, , \quad i,j=1, \dots, N \, ,
	\label{eq:demkovfradkinDIII}
\end{equation}
which satisfy the following relations 
\begin{equation}
\hat{H}^{(\lambda)}=\frac{1}{2}\sum_{i=1}^N 	\hat{\mathcal{F}}_{ii}^{(\lambda)} \, , \quad	\big[\hat{H}^{(\lambda)}, \hat{\mathcal{F}}_{ij}^{(\lambda)}  \big]=0\, ,\quad  i,j=1, \dots, N.
	\label{eq:commrel}
\end{equation} 
Therefore, $\hat{H}^{(\lambda)}$  adequately represents a generalization by reflections of the quantum Darboux III oscillator \cite{BEHRR2011}.  
  
\begin{rem}
This model was recently introduced in \cite{BNPHD2024}, where the corresponding spectral problem was solved. It is worth underlining that   the previous work   also showed that, due to the Dunkl extension and the fact that the model is defined in a space with non-constant curvature, the Hilbert space and its corresponding inner product must be appropriately defined, as expected.
\end{rem}


\subsection{Generalized Dunkl-Darboux III system} 
\label{subsec4.6}
\noindent 
We preserve the function (\ref{eq:quantumDIII}), with the same underlying space $\bD^N$ (\ref{spaceD}) and oscillator potential, but now consider the complete  realization of   $\mathfrak{sl}(2, \mathbb{R})$  with $\{\beta_i , \gamma_i \}_{i=1}^N\ne 0$  \eqref{NDrepquantum},
obtaining the Hamiltonian
\begin{align}
\hat{H}^{(\lambda)}&= \frac{1}{2(1+\lambda \hbx^2)}\left(\hbpi^2+\sum_{i=1}^N \frac{\beta_i+\gamma_i \hat{R}_i}{\hat{x}_i^2}\right)+\frac{\omega^2}{2}\frac{\hbx^2}{1+\lambda \hbx^2}\nonumber \\[2pt]
& = \frac{1}{2(1+\lambda \bx^2)}\left(-\hbar^2\Delta_\D+\sum_{i=1}^N \frac{\beta_i+\gamma_i \hat{R}_i}{x_i^2}\right)+\frac{\omega^2}{2}\frac{\bx^2}{1+\lambda \bx^2} \, ,
\label{eq:gDIII}
\end{align}
so that the left and right quantum integrals are given by \eqref{left1}. Moreover, the Hamiltonian is still MS as it is endowed with the following  $N$ additional  quantum integrals 
\begin{equation}
	\hat{\mathcal{F}}_{i}^{(\lambda)} = \hpi_i^2+\frac{\beta_i+\gamma_i \hat{R}_i}{\hat{x}_i^2}+ \hx_i^2\left(\omega^2 -2 \lambda \hat{H}^{(\lambda)}\right)\, , \quad i=1, \dots, N \, ,
	\label{DIIIgen}
\end{equation}
  fulfilling
\begin{equation}
\big[\hat{H}^{(\lambda)}, 	\hat{\mathcal{F}}_{i}^{(\lambda)}  \big]=0\, ,\quad  i=1, \dots, N\, .
\label{eq:commrel2}
\end{equation}
Let us  notice that the Hamiltonian (\ref{eq:gDIII}) can be expressed as
\begin{equation}
\hat{H}^{(\lambda)}=\frac{1}{2}\sum_{i=1}^N 	\hat{\mathcal{F}}_{i}^{(\lambda)}\, .
\label{eq:DFH}
\end{equation}
Clearly, the limit $\lambda=0$ leads to the (flat) Dunkl-Smorodinsky-Winternitz system of Section~\ref{subsec4.2}.  Also note that if we drop the non-central terms,  $\{\beta_i , \gamma_i \}_{i=1}^N= 0$, the $N$ constants (\ref{DIIIgen}) collapse to the diagonal components of the curved Dunkl-Demkov-Fradkin tensor \eqref{eq:demkovfradkinDIII}.

Consequently, the model here presented provides an extension by reflections of the quantum analogue of the generalized Darboux III oscillator introduced in \cite{BEHR08a}, which to our knowledge has not been considered so far  in the literature. 

\section{Maximally superintegrable Dunkl-Kepler-Coulomb-type systems}
\label{sec5}
\noindent Following the same steps as described at the beginning of Section~\ref{sec4}, we construct six MS systems  of Dunkl-KC-type  in a  ``parallel" way to those given for the Dunkl oscillator-type in Sections~\ref{subsec4.1}--\ref{subsec4.6}. This means that instead of deriving a Dunkl-Demkov-Fradkin tensor, for each model we present a Dunkl-LRL  type $N$-vector  or  one quantum integral quadratic in the momenta which arises as a generalization of its $N$-th component. It should be stressed that,  with the exception of the well-known Dunkl-KC system summarized in Section~\ref{subsec5.1},  the  five  MS systems presented in  Sections~\ref{subsec5.2}--\ref{subsec5.6} are actually new models.


\subsection{Dunkl-KC system}
\label{subsec5.1}

\noindent   This model arises with the following choice of the function (\ref{eq:ham}):
\begin{equation}
F\big(\hat J_+,\hat J_-,\hat J_3\big)= \frac{\hat J_+}{2} -\frac{k}{ {\hat J_-^{1/2}}} \, \quad (k \in \mathbb{R}) 
\label{eq:kep-coul}
\end{equation}
under the   $N$D differential-difference realization \eqref{NDrepquantumdunkl}. This corresponds to taking $V\big(\hat J_-\big)=- k/ {\hat J_-}^{1/2}$ in \eqref{eq:Euclideanj} (so $V_{\textsf{KC}}$ (\ref{HOKC})). The Hamiltonian operator reads
	\begin{equation}
\hat{H} = \frac{\hbpi^2}{2}- \frac{k}{|\hbx|} = -\frac{\hbar^2}{2}\Delta_\D-\frac{k}{|\bx|} \, .
\label{eq:dunklkepler}
\end{equation}
Also in this case, since the realization is the same as the Dunkl oscillator in Section~\ref{subsec4.1}, the left and right quantum
integrals are the ones reported in \eqref{dunklang}. The resulting  Dunkl-KC system turns out to be MS as there exists a Dunkl extension of the LRL vector \cite{G1975, G1976}. In the $N$D case, its components read:
\begin{equation}
\hat{\mathcal{A}}_i = \frac{1}{2} \sum_{j=1}^N\big\{\hat{\Lambda}_{ij}, \hpi_j\big\}-k\frac{\hx_i}{|\hbx|}+\imath \hbar \mu_i \hr_i \hpi_i \, , \quad i = 1, \dots, N \, ,
\label{eq:dunkLRL}
\end{equation}
where  $\{\cdot, \cdot\}$ indicates the anticommutator. Each component of this Dunkl-LRL $N$-vector commutes with the Hamiltonian operator \eqref{eq:dunklkepler}:
\begin{equation}
	\big[\hat{H}, \hat{\mathcal{A}}_i \big]=0\, , \quad i=1, \dots, N
\label{commdDLRL}
\end{equation}
as it can  be  directly checked. Similarly to the standard (reflectionless) case, we find the following functional relation:
\begin{equation}
 \sum_{i=1}^N \hat{\mathcal{A}}_i^2 = 2 \hat{H}\left(\hat{\boldsymbol{\Lambda}}^2 +\hbar^2 \left(\sum_{i=1}^N \mu_i \hr_i+\frac{N-1}{2}\right)^2\right)+k^2\, .
\label{funrel}
\end{equation}

\begin{rem}
For $N = 2$, we recover the model studied in \cite{GLV2015}, whereas for $N = 3$ the one appearing in \cite{GSAE2019,GS2020}. The $N$D case can be found in \cite{Ghaz2021, Quesne2024}. We also mention \cite{FH2022}, where a Dunkl version of the LRL vector associated with an arbitrary finite Coxeter group is introduced.
\end{rem}


\subsection{Quasi-generalized Dunkl-KC system}
	\label{subsec5.2}

\noindent As we have already mentioned in Section~\ref{sec3}, it is well-known that when all the $N$ $\beta_i$-terms are added to the standard KC potential,  the resulting generalized KC system is MS but endowed with  integrals  of LRL type quartic in the momenta \cite{VE2008,KC2009,TD2011}.
Nevertheless, when one parameter $\beta_i$ vanishes,    a quadratic quantum integral arises and the system is the so-called quasi-generalized  KC system.
Therefore, here we  restrict ourselves to constructing the $N$D quasi-generalized Dunkl-KC system, which represents an extension by reflections of the quantum quasi-generalized KC system \cite{Evans90b,KWMP1999, KWMP2002, RW2002, LMZ2018}. Such model is provided by the same  choice of the function $F$ (\ref{eq:kep-coul})  but
 expressed  through the $N$D realization of the $\mathfrak{sl}(2, \mathbb{R})$ \eqref{NDrepquantum} with $\beta_N = \gamma_N = 0$.   The Hamiltonian results in:
\begin{equation}
	\hat{H} = \frac{\hbpi^2}{2}+\sum_{j=1}^{N-1} \frac{\beta_j +\gamma_j \hr_j}{2\hx_j^2}-\frac{k}{|\hbx|}= -\frac{\hbar^2}{2}\Delta_\D+\sum_{j=1}^{N-1} \frac{\beta_j +\gamma_j \hr_j}{2x_j^2}-\frac{k}{|\bx|} \, .
	\label{eq:qgh}
\end{equation}
In this case, the left and right quantum integrals are given by \eqref{left1}  with $\beta_N = \gamma_N = 0$. This $N$D quantum model of Dunkl type turns out to be MS due to the existence of the following (second-order) quantum integral:
\begin{equation}
	\hat{\mathcal{A}}_N= \sum_{j=1}^{N-1} \left(\frac{1}{2} \big\{\hat{\Lambda}_{Nj}, \hpi_j\big\}+\big(\beta_j +\gamma_j \hr_j\big)\frac{\hx_N}{\hx_j^2}\right)-k\frac{\hx_N}{|\hbx|}+\imath \hbar \mu_N \hr_N \hpi_N \, , 
	\label{eq:dunkqglRL}
\end{equation}
which generalizes the $N$-th component of the Dunkl-LRL $N$-vector \eqref{eq:dunkLRL}. In fact, it can be checked by direct computations that 
\begin{equation}
\big[\hat{H},\hat{\mathcal{A}}_N \big]=0 \, .
\end{equation}

\begin{rem}
	If we would have fixed $\beta_{l} =\gamma_{l}= 0$ for a given index $l$ among $\{1,\dots,N\}$ in \eqref{eq:qgh} we would have obtained the conservation of a generalized version of the $l$-th component of the Dunkl LRL vector. 
\end{rem}


 \subsection{Dunkl-KC system on the sphere and hyperbolic space}
\label{subsec5.3}

\noindent   
We now aim to construct the corresponding MS Dunkl-KC system on  the three spaces $\mani^N_\kappa$ (\ref{zf}) in terms of the curvature $\kappa$, thus covering $\bS^N$ and $\bH^N$, together with their flat contraction giving rise to the proper Dunkl-KC system in Section~\ref{subsec5.1}. 
As in Section~\ref{subsec4.3}, we  again use the Poincar\'e projective coordinates (\ref{ze}).
The  free quantum Hamiltonian is thus given by (\ref{TTk}), while the KC potential on $\mani^N_\kappa$   reads as~\cite{BH2007,annals2009}
\be
V^{(\kappa)}_{\textsf{KC}} = -k\frac{1-\kappa \hat J_-}{ {\hat J_-^{1/2}}} =- k\frac{1-\kappa \hbx^2}{|\hbx|}  \, \quad (k, \kappa \in \mathbb{R}) \, ,\\
 \label{z53a}
\ee
leading to the Euclidean $V_{\textsf{KC}}$ (\ref{HOKC}) for $\kappa=0$. The curved KC potential 
comes from its expression  in terms of  the $N+1$ ambient coordinates $(\amb_0,\ambb)$  (\ref{constraint}) and then substituting the Poincar\'e coordinates in the form
\be
V^{(\kappa)}_{\textsf{KC}} = - \tilde k\frac{\amb_0}{|\ambb|}= - \tilde k  {\frac{\amb_0}{\sqrt{\big(1-\amb_0^2\big)/\kappa}}}=-k \frac{  1-\kappa \bx^2     }{|\bx|}\, , \quad {\tilde k}  = 2k\, .
\label{z53b}
\ee  
By introducing the geodesic radial coordinate $r$ (\ref{geod}), from the particle to the origin along the geodesic $\ell$,  we find a more common
expression for $V^{(\kappa)}_{\textsf{KC}} $, namely (see, e.g., \cite{carJMP2005} and references therein)
\be
\begin{aligned}
\bS^N \ (\kappa=+1)\!:&  \quad V^{(+)}_{\textsf{KC}} = -  \frac{\tilde k}{\tan  r}\,  ,\quad\  \,   \amb_0= \cos r  \, ,\quad\ \      |\bx|= \tan \frac { r}2\, \\[2pt]
\bH^N \ (\kappa=-1)\!:&\quad  V^{(-)}_{\textsf{KC}} = -  \frac{\tilde k}{\tanh  r} \,  ,\quad  \amb_0= \cosh r  \, ,\quad  |\bx|= \tanh \frac { r}2\, ,
\end{aligned}
\label{KCSS}
\ee
to be compared with  $V^{(\kappa)}_{\textsf{HO}}$  in (\ref{HiggsSS}); observe that, in fact, 
$V^{(\kappa)}_{\textsf{HO}} \propto \big( V^{(\kappa)}_{\textsf{KC}} \big)^{-2}$.

 Consequently, the   Dunkl-KC system on the spaces of constant curvature $\mani^N_\kappa$ is obtained through  the following choice of the function $F$:
\begin{equation}
	F\big(\hat J_+,\hat  J_-,\hat  J_3\big )=\big(1+\kappa \hat J_-\big)^2 \frac{\hat  J_+}{2} -k\frac{1-\kappa \hat  J_-}{ {\hat  J_-^{1/2}}} \, \quad (k,\kappa \in \mathbb{R}) \, ,
	\label{eq:curvedkep-coul}
\end{equation}
 and  then introducing   the differential-difference realization \eqref{NDrepquantumdunkl},
 finding the   curved  spherical/hyperbolic Dunkl-KC Hamiltonian operator  given by
\begin{equation}
 \hat{H}^{(\kappa)} =\big(1+\kappa \hbx^2\big)^2 \frac{\hbpi^2}{2}- k\frac{1-\kappa \hbx^2}{|\hbx|} = -\frac{\hbar^2(1+\kappa \bx^2)^2 }{2}\Delta_\D-k\frac{1-\kappa \bx^2 }{|\bx|} \, .
 \label{eq:dunkcurvedlkepler}
\end{equation}
Hence, the left and right quantum
integrals are  those written in \eqref{dunklang}. The MS property of $ \hat{H}^{(\kappa)}$ is established 
by   a Dunkl extension of the curved LRL $N$-vector  \cite{BH2007}, whose components
 turn out to be
\begin{align}
	\hat{\mathcal{A}}_i^{(\kappa)}&= \frac{1}{2} \sum_{j=1}^N \left\{\hat{\Lambda}_{ij}, \hat{\Gamma}^{(\kappa)}_j\right\}-k\frac{\hx_i}{|\hbx|}+\imath \hbar \mu_i \hr_i \hat{\Gamma}^{(\kappa)}_i\nonumber\\[2pt]
	&\qquad + \imath \hbar  \kappa (N-2)  \left(  \hbx^2 \hpi_i - \hat{x}_i (\hbx \cdot \hbpi)+\imath \hbar \hat{x}_i\left(\sum_{j=1}^N \mu_j \hat{R}_j+\frac{N-1}{2} \right)\right) \, ,
	\label{eq:curvedDunkLRL}
\end{align}
where $\hat{\Gamma}^{(\kappa)}_i$ is the curved Dunkl momentum operator  defined in \eqref{eq:operator}. Each component of this curved Dunkl-LRL $N$-vector commutes with the Hamiltonian operator \eqref{eq:dunkcurvedlkepler}, i.e.:
\begin{equation}
	\big[\hat{H}^{(\kappa)}, \hat{\mathcal{A}}^{(\kappa)}_i \big]=0\, , \quad i=1, \dots, N
	\label{commdcurvedDLRL}
\end{equation}
as it can be directly checked. Clearly, in the flat $\kappa \to 0$ limit, the Dunkl-LRL $N$-vector \eqref{eq:dunkLRL} is recovered (recall that $\hat{\Gamma}^{(\kappa)}_i\to \hpi_i$ when $\kappa\to 0$).

Therefore,  this novel model  represents an extension by reflections of the quantum KC system on the $N$D sphere and hyperbolic space.


\subsection{Quasi-generalized Dunkl-KC system on the sphere and hyperbolic space}
\label{subsec5.4}
\noindent 
The generalized KC on $\bS^N$ and $\bH^N$ is known to be quadratically MS whenever, at least, one $\beta_i$-potential vanishes \cite{KC2009} as   on  $\bE^N$ \cite{VE2008,TD2011}.
 Thus, we take the same function   $F$ (\ref{eq:curvedkep-coul})  and introduce the   $N$D realization of   $\mathfrak{sl}(2, \mathbb{R})$ \eqref{NDrepquantum} with $\beta_N = \gamma_N = 0$, similarly to Section~\ref{subsec5.2}. The Hamiltonian operator results in:
\begin{align}
	\hat{H}^{(\kappa)} &= \frac{\big(1+\kappa \hbx^2\big)^2}{2}\left(\hbpi^2+\sum_{j=1}^{N-1} \frac{\beta_j +\gamma_j \hr_j}{\hx_j^2}\right)-k\frac{1-\kappa \hbx^2}{|\hbx|} \nonumber\\[2pt]
 &= \frac{ (1+\kappa \bx^2)^2}{2}\left(-\hbar^2\Delta_\D+\sum_{j=1}^{N-1} \frac{\beta_j +\gamma_j \hr_j}{x_j^2}\right)-k\frac{1-\kappa \bx^2}{|\bx|} \, ,
\label{eq:qghc}
\end{align}
 and the left and right quantum integrals are given by \eqref{left1} with $\beta_N = \gamma_N = 0$.
An additional quantum integral is found to be
\begin{align}
	\hat{\mathcal{A}}^{(\kappa)}_N=& \sum_{j=1}^{N-1} \left(\frac{1}{2} \left\{\hat{\Lambda}_{Nj}, \hat{\Gamma}^{(\kappa)}_j\right\}+\big(1-\kappa \hbx^2 \big) \big(\beta_j +\gamma_j \hr_j \big)\frac{\hx_N}{\hx_j^2}\right)-k\frac{\hx_N}{|\hbx|}+\imath \hbar \mu_N \hr_N \hat{\Gamma}^{(\kappa)}_N  \nonumber \\
	&+ \imath \hbar  \kappa (N-2)  \left(  \hbx^2 \hpi_N - \hat{x}_N (\hbx \cdot \hbpi)+\imath \hbar \hat{x}_N\left(\sum_{j=1}^N \mu_j \hat{R}_j+\frac{N-1}{2} \right)\right) \, ,
	\label{eq:dunkqcurvedglRL}
\end{align}
which proves  the MS property of $\hat{H}^{(\kappa)}$, since it can be directly verified that 
\begin{equation}
	\big[\hat{H}^{(\kappa)},\hat{\mathcal{A}}^{(\kappa)}_N \big]=0 \, .
\end{equation}
Hence, $\hat{\mathcal{A}}^{(\kappa)}_N$ turns out to be  the generalization of the  $N$-th component of the curved Dunkl-LRL $N$-vector \eqref{eq:curvedDunkLRL} in the presence of non-central terms with $\{\beta_j, \gamma_j\}_{j=1}^{N-1}\ne  0$. 
 Again, the operators $\hat{\Gamma}^{(\kappa)}_j$ are  defined by \eqref{eq:operator}. Obviously, in the flat $\kappa \to 0$ limit, the quantum constant \eqref{eq:dunkqcurvedglRL} reduces to   \eqref{eq:dunkqglRL} in Section~\ref{subsec5.2}.
 
 Therefore, we have deduced the quasi-generalized Dunkl-KC system on the $N$D sphere and hyperbolic space, which is the counterpart with reflections of the quasi-generalized KC system on such spaces introduced in \cite{KC2009}.

 In addition, it is worth remarking that an  interpretation of the $N-1$  $\beta_j$-terms as non-central oscillators can be  performed straightforwardly on $\bS^N$, in exactly  the same way as in Section~\ref{subsec4.4} for  the generalized spherical Dunkl oscillator. In this case, let us write the Hamiltonian $\hat{H}^{(\kappa)} $ as the sum of the kinetic operator $ \hat{\TT}^{(\kappa)}$  (\ref{TTk})
 and the quasi-generalized curved Dunkl-KC potential  
 \be
  \hat{V}_{\textsf{qgKC}}^{(\kappa)}= \hat{V}_{\textsf{KC}}^{(\kappa)}+ \frac{(1+\kappa \bx^2)^2}{2} \sum_{j=1}^{N-1} \frac{\beta_j+\gamma_j \hat{R}_j}{x_j^2} \, ,
   \label{gKCk}
 \ee
where $\hat{V}_{\textsf{KC}}^{(\kappa)}$ is the curved KC potential given in (\ref{z53a}). 
  If we consider the same $N-1$ fixed points $O_j$ $(j=1,\dots, N-1)$,  positive parameters $\beta_j=\omega_j^2/4>0$ and distances $r_j$ used in the expression   (\ref{gHOSN}), 
then we obtain that  $\hat{V}_{\textsf{qgKC}}^{(\kappa)}$ on $\bS^N$, with $\kappa=+1$, becomes
 \be
   V^{(+)}_{\textsf{qgKC}} =- \frac{  \tilde k}{\tan  r}+\frac 12 \sum_{j=1}^{N-1}   \omega_j^2  \tan^2\! r_j  +2\sum_{j=1}^{N-1} \gamma_j \big(1+\tan^2\! r_j\big){ \hat{R}_j} +\frac{1}{2} \sum_{j=1}^{N-1} \omega_j^2 \, ,
      \label{gKCk2}
 \ee
  which, in analogy with    (\ref{gHOSN}), shows the superposition of the spherical KC potential with centre at the origin $O$  with $N-1$ non-central spherical oscillators with centres at the points $O_j$, along with  $N-1$ spherical non-central oscillator-reflection potentials.


\subsection{Dunkl Taub-NUT system}
\label{subsec5.5}

\noindent 
As mentioned in Section~\ref{subsec4.5}, the Darboux III oscillator can be seen as the quadratic MS system closest to the isotropic oscillator on a space of non-constant curvature. As far as the KC potential is concerned, this role is played by the so-called Taub-NUT system~\cite{BEHRR2014} (see \cite{LatiniNUT} for a complete description and solution of the classical system). The underlying Taub-NUT space  appears  via the conformal function  $\ff(|\bx|)= {(1+\eta/ |\bx|)^{1/2}}$, such that  the   metric (\ref{zb}) and scalar curvature  (\ref{zc}) read
\be
\text{d} \metric^2=  \left(1+\frac{\eta}{|\bx|} \right)\text{d}\bx^2\, ,\quad \RR(|\bx|)=   \eta (N-1)\,\frac{4(N-3)|\bx|+ 3 \eta(N-2)  }{4|\bx|(\eta+ |\bx|)^3}\, , \quad  |\bx|\ne 0,
\label{tna}
 \ee
where the parameter $\eta\in \bR$. Therefore, if $\eta>0$  the coordinates   $\bx\in (-\infty,+\infty)$, but 
 when  $\eta=-|\eta|<0$ the domain is restricted to the exterior of the ball $|\bx| > |\eta|$.
 
Then, to construct  the $N$D Dunkl Taub-NUT system we consider the quantum kinetic term coming from the metric (\ref{tna}), applying  the direct Schr\"odinger quantization (\ref{zi}), and add the quantum Taub-NUT potential introduced in \cite{BEHRR2014}. This procedure corresponds to  take the following function $F$
\begin{equation}
 F\big(\hat J_+,\hat J_-,\hat J_3 \big)= \frac{ {\hat J_-^{1/2}}}{\eta+ {\hat J_-^{1/2}}}\frac{\hat J_+}{2} -\frac{k}{\eta+ {\hat J_-^{1/2}}}  \quad (k, \eta \in \mathbb{R}) \, ,
	\label{eq:quantumTN}
\end{equation}
giving rise, via  the realization  \eqref{NDrepquantumdunkl}, to the    Dunkl Taub-NUT Hamiltonian
\begin{equation}
	\hat{H}^{(\eta)} = \frac{|\hbx|}{\eta+|\hbx|}\frac{\hbpi^2}{2}-\frac{k}{\eta+ |\hbx|} = -\frac{\hbar^2 |\bx|}{2(\eta+ |\bx|)}\Delta_\D-\frac{k}{\eta+|\bx|} \, .
	\label{eq:DTN}
\end{equation}
This system provides an extension  by reflections of the quantum Taub-NUT system \cite{BEHRR2014}. 
 On the other hand,  $\hat{H}^{(\eta)}$ can also be  interpreted as a one-parameter MS deformation of the Dunkl-KC system in Section~\ref{subsec5.1}, now defined on the space of non-constant curvature (\ref{tna}). 
In particular,  the left and right quantum integrals are once more given by \eqref{dunklang}, moreover, the model turns out to be MS since there exists  a deformed version of the Dunkl-LRL $N$-vector \eqref{eq:dunkLRL}, whose components read:
\begin{equation}
	\hat{\mathcal{A}}_i^{(\eta)} = \frac{1}{2} \sum_{j=1}^N\big\{\hat{\Lambda}_{ij}, \hpi_j\big\}-\frac{\hx_i}{|\hbx|}\left(k+\eta \hat{H}^{(\eta)}\right)+\imath \hbar \mu_i \hr_i \hpi_i \, , \quad i = 1, \dots, N\, .
	\label{eq:DunkLRL}
\end{equation}
By direct computations, it can be readily verified that:
\begin{equation}
	\big[\hat{H}^{(\eta)},\hat{\mathcal{A}}_i^{(\eta)} \big]=0 \, , \quad i=1, \dots, N \,.
	\label{commHA}
\end{equation}
In this $\eta$-deformed case, the functional relations \eqref{funrel} extends to:
\begin{equation}
	\sum_{i=1}^N \left(\hat{\mathcal{A}}_i^{(\eta)}\right)^2 = 2 \hat{H}^{(\eta)}\left(\hat{\boldsymbol{\Lambda}}^2 +\hbar^2 \left(\sum_{i=1}^N \mu_i \hr_i+\frac{N-1}{2}\right)^2\right) +\left(k+\eta \hat{H}^{(\eta)}\right)^2\, .
	\label{funreldef}
\end{equation}
Let us conclude by briefly commenting that for this model a natural Hilbert space candidate would be: 
\begin{equation}
	L^2\left(\mathbb{R}^N,\left(1+\frac{\eta}{|\boldsymbol{x}|}\right) \prod_{i=1}^{N}|x_i|^{2\mu_i}\text{d}\bx\right), 
	\label{HSTN}
\end{equation}
with corresponding inner product:
\begin{equation}
	\braket{\gga | \ggb}_{\boldsymbol{\mu},\eta}:=	\int_{\mathbb{R}^N} \overline{\gga(\bx)}\ggb(\bx)\left(1+\frac{\eta}{|\boldsymbol{x}|}\right) \prod_{i=1}^{N}|x_i|^{2\mu_i}\text{d}\bx \, ,
	\label{eq:innprodTN}
\end{equation}
which  is consistent with (\ref{eq:inner}) and reduces to the one given in~\cite{BEHRR2014} when all $\mu_i=0$.
  

\subsection{Quasi-generalized Dunkl Taub-NUT system}
\label{subsec5.6}

\noindent 
In contrast to the generalized Dunkl-Darboux III oscillator in Section~\ref{subsec4.6}, for which $N$  
 generalized  diagonal components (\ref{DIIIgen}) of the $\lambda$-deformed Demkov-Fradkin type tensor (\ref{eq:demkovfradkinDIII})   were previously known for the classical system \cite{BEHR08a}, to our knowledge, there is no yet such result for the Taub-NUT system with centrifugal terms, although the generalized classical system was introduced in \cite{annals2009}, but without computing its additional integrals.
As our last application, let us construct the quasi-generalized Dunkl Taub-NUT system with $N-1$ $\beta_j$- and 
   $\gamma_j$-potentials $(j=1,\dots,N-1)$ which, as a byproduct, will provide the MS property for the lacking standard quantum system.

With this aim, we keep the  function $F$ (\ref{eq:quantumTN}) and introduce   the $N$D realization of the $\mathfrak{sl}(2, \mathbb{R})$ \eqref{NDrepquantum} with $\beta_N = \gamma_N = 0$. The resulting Hamiltonian   turns out to be
\begin{align}
 \hat{H}^{(\eta)} &= \frac{|\hbx|}{2(\eta+|\hbx|)} \left(\hbpi^2+\sum_{j=1}^{N-1} \frac{\beta_j +\gamma_j \hr_j}{\hx_j^2}\right)-\frac{k}{\eta+ |\hbx|} \\[2pt]
 & = \frac{ |\bx|}{2(\eta+ |\bx|)}\left(-\hbar^2\Delta_\D +\sum_{j=1}^{N-1} \frac{\beta_j +\gamma_j \hr_j}{x_j^2}	\right)-\frac{k}{\eta+|\bx|} \, .
 \label{eq:qgDTN}
\end{align}
The left and right quantum integrals are given by \eqref{left1} with $\beta_N = \gamma_N = 0$. Furthermore, we find an additional quadratic   quantum integral given by
\begin{equation}
	\hat{\mathcal{A}}^{(\eta)}_N= \sum_{j=1}^{N-1} \left(\frac{1}{2} \big\{\hat{\Lambda}_{Nj}, \hpi_j \big\}+\big(\beta_j +\gamma_j \hr_j \big)\frac{\hx_N}{\hx_j^2}\right)-\frac{\hx_N}{|\hbx|}\left(k+\eta \hat{H}^{(\eta)}\right)+\imath \hbar \mu_N \hr_N \hpi_N \, , 
	\label{eq:dunkqgLRL}
\end{equation}
which thus generalizes the $N$-th component of the $\eta$-deformed Dunkl LRL vector \eqref{eq:DunkLRL}. By direct computations, it can be shown that:
\begin{equation}
	\big[\hat{H}^{(\eta)},\hat{\mathcal{A}}_N^{(\eta)} \big]=0 \, .
	\label{commHA2}
\end{equation}
Clearly, in the $\eta \to 0$ limit,  the quantum integral reduces to \eqref{eq:dunkqglRL}.


\section{Concluding remarks and open perspectives}
\label{sec6}
\noindent Dunkl superintegrable systems associated with the $\mathbb{Z}_2^N$
reflection group (for $N=2,3$ and also for general $N$) have attracted significant attention in recent years. In this paper, we have proved that these quantum Hamiltonians can all be understood as possessing the same underlying $\mathfrak{sl}(2,\mathbb{R})$ coalgebra symmetry,   which has remained hidden until now. The link between the coalgebra symmetry approach to superintegrability and Dunkl superintegrable systems is made possible by the existence of a differential-difference realization of the aforementioned Lie algebra. This approach, which has proven to be highly useful in many areas of (super)integrability, both continuous and discrete, is therefore also applicable in the Dunkl realm. This application was previously absent in the literature. As a result, we have introduced a novel family of 
$N$D Dunkl Hamiltonians that are QMS by construction: the $2N-3$ (second-order) Dunkl quantum integrals are provided by the left and right Casimirs of the coalgebra. Several MS subcases of this family are already known in the literature, including the well-known Dunkl oscillator and Dunkl-KC system. Thanks to the coalgebra symmetry, new applications have also been  provided, including models defined on the constant curvature spaces $\bS^N$ and $\bH^N$, as well as on non-constant curvature spaces.  In addition, MS generalizations of the Euclidean and curved Dunkl oscillators and KC systems have been constructed through the introduction of   non-central (centrifugal-reflection) potentials, which for both the Dunkl oscillator  and KC system on  $\bS^N$  have been interpreted as non-central Higgs oscillators. It is worth remarking that in all specific applications, the MS property has been established by explicitly presenting, at least, one additional quantum quadratic integral. Namely, for each model, we obtained a Dunkl-Demkov-Fradkin type tensor, or $N$ quantum integrals extending its diagonal components, in the case of Dunkl oscillator-type Hamiltonians, while for Dunkl KC-type systems, we obtained a Dunkl-LRL $N$-vector, or an extension of one of its components (the $N$-th component).

These models, of course, do not exhaust the set of MS systems of Dunkl type that can be introduced through coalgebra symmetry and further explored in future works. If we keep the Hamiltonian function  $\hat H=F\big(\hat J_+,\hat  J_- \big)$, which indeed covers the  $6+6$ specific MS Dunkl oscillator  and KC-type  systems  discussed here, we find several natural possibilities worth studying:

\begin{itemize}

\item For flat and curved Dunkl KC-type systems, only up to $N-1$ non-central terms have been considered  due to  the quadratic MS condition, giving rise to their quasi-generalized counterparts. From the  already known results for generalized KC Hamiltonians on spaces of constant curvature \cite{VE2008,KC2009,TD2011}, one can expect that there exist $N$   quartic quantum integrals of  Dunkl-LRL type for any generalized Dunkl-KC system. These quantum integrals, in the flat Euclidean case, should appear as a Dunkl generalization of the ones reported in \cite{LMZ2021}.

 \item   Classical MS systems with a central potential on spherically symmetric spaces (\ref{zb}) have  been  classified for $N=3$ in~\cite{Bertrand2008} from  previous results presented in \cite{Perlick}. These are called  Bertrand Hamiltonians and are found to be  of oscillator or KC type, and  generally  endowed 
  with higher-order constants of motion.  Their corresponding Dunkl versions could be derived from the family (\ref{eq:curvedaddterms})  by using appropriate expressions for the conformal function and central potential. It is worth noting that  the  classical MS systems  in this work  are, in fact, derived from  3D Bertrand Hamiltonians. Nevertheless, obtaining explicit quantum integrals for such Dunkl-Bertrand Hamiltonians is not a trivial task at all.
 
 \item Following the results in \cite{annals2009} (see references therein), a  Dirac monopole-type term  could be added in the Dunkl context to the family of (curved)  Hamiltonians  (\ref{eq:curvedaddterms}) as
 \be
 \hat{H} = \frac { 1 }{ 2 \ff  \big(  {\hat{J}_-^{1/2} } \big)^2 }\left( \hat{J}_+ + \frac{\varrho^2}{\hat{J}_-}  \right) +  V\!\left(\!\sqrt{\hat{J}_-} \,\right) \,  
 \ee
 where $\varrho$ is a real constant, thus opening  lines of research for the study of  the Dunkl extension of physically relevant systems such as the  (curved) multifold Kepler and MIC-Kepler Hamiltonians~\cite{IwaiKatayama}.
\end{itemize}
 
\noindent Clearly, other systems can be derived from the $\mathfrak{sl}(2,\mathbb{R})$ coalgebra symmetry, but not belonging  to the subfamily  $\hat H=F\big(\hat J_+,\hat  J_- \big)$. Let us  mention, for example, the Zernike Hamiltonian \cite{Zernike}, which is a 2D quantum system that has attracted significant attention in recent years  \cite{PWY2017, PSAWY2017, PWY2017b, Fordy2018}, particularly in relation to its higher-order (momentum-dependent) generalizations in both classical mechanics \cite{BGSH2023, GGK2024} and quantum mechanics \cite{Campoamor-Stursberg:2025jke}. It was shown in \cite{BGSH2023} that while its classical version can be understood as possessing an $\mathfrak{sl}(2,\mathbb{R})$ coalgebra symmetry, its quantum version corresponds to having an underlying (co)algebra which is isomorphic to $\mathfrak{gl}(2,\mathbb{R})$.  The complete Dunkl extension of the generalized quantum Zernike Hamiltonian would be given by
\be
\hat H= F\big(\hat J_+^{[2]},\hat  J_3^{[2]} \big)= \hat J_+^{[2]} +\sum_{l=1}^n\zeta_l \left( \hat J_3^{[2]} \right)^n
\ee
expressed under the   realization  (\ref{NDrepquantum}) for $N=2$  and where $\zeta_l$ are arbitrary real or pure imaginary constants.

Another interesting direction for further exploration would be to investigate the role of Dunkl operators within this coalgebra framework. As a further potential future direction, we observe that determining the polynomial algebra generated by the $2N-3$ left and right Dunkl-type quantum integrals would extend the quadratic structures presented in \cite{Latini2019} to include reflections. From this perspective, let us recall that the higher-rank Racah algebra $R(N)$ was obtained in \cite{DBGVV2018} as the symmetry algebra of the Laplace-Dunkl operator associated to the $\mathbb{Z}^N_2$ reflection group. 
Finally,  the study of the symmetry algebra of specific models could allow an algebraic derivation of the spectrum, providing some hints about the expected rigorous solution of the spectral problem by solving the corresponding Schr\"odinger equation.

    In conclusion, the main message of this work is that the coalgebra symmetry method for superintegrable systems also finds a natural application when Dunkl operators are considered, opening the way for the introduction of many new superintegrable systems of Dunkl type.


\section*{Acknowledgements}

\phantomsection
\addcontentsline{toc}{section}{Acknowledgements}

\noindent F.J.H. has been partially supported by Agencia Estatal de Investigaci\'on (Spain) under  the grant PID2023-148373NB-I00 funded by MCIN/AEI/10.13039/501100011033/FEDER, UE.  F.J.H. also acknowledges support  by the  Q-CAYLE Project  funded by the Regional Government of Castilla y Le\'on (Junta de Castilla y Le\'on, Spain) and by the Spanish Ministry of Science and Innovation (MCIN) through the European Union funds NextGenerationEU (PRTR C17.I1). The research of D.L.~has been partially funded by MUR - Dipartimento di Eccellenza 2023-2027, codice CUP G43C22004580005 - codice progetto   DECC23$\_$012$\_$DIP and partially supported by INFN-CSN4 (Commissione Scientifica Nazionale 4 - Fisica Teorica), MMNLP project. D.L. is a member of GNFM, INdAM.


\appendix
\setcounter{subsection}{0}
\section{Algebraic structure generated by curved Dunkl momenta on $\bS^N$  and   $\bH^N$ with Dunkl angular momenta and reflection operators}
\label{AA}

 \noindent 
 Let us consider the set of   $N(N-1)/2+N+N=N(N+3)/2$ quantum operators   $\big\{\hat{\Lambda}_{ij} , \hat{R}_i, \hat{\Gamma}_i^{(\kappa)}\big\}$ with $1\le i<j\le   N$,  formed by  the $N(N-1)/2$ Dunkl angular momenta (\ref{eq:angmomdunkl}), $N$ reflections (\ref{reflections}) along with the   $N$ curved  Dunkl momenta (\ref{eq:operator}). For the sake of completeness, we recall  their explicit realization:
 \be
\begin{aligned}
\hat{\Lambda}_{ij} &   = \hx_i \hpi_j - \hx_j \hpi_i \, ,  \quad \hr_i\Psi(\bx) =\Psi(\sigma_i (\bx))  \\[2pt]
\hat{\Gamma}_i^{(\kappa)}&=\big(1-\kappa \hbx^2\big)\hpi_i+2 \kappa  \hat{x}_i \left((\hbx \cdot \hbpi)-\imath \hbar \sum_{j=1 }^N \mu_j \hat{R}_j\right)\, ,
 \end{aligned}
 \label{A1}
\ee
where  $\kappa$ is the curvature of $N$D sphere $\bS^N$ and hyperbolic space  $\bH^N$. It can be shown that these  $N(N+3)/2$ operators close on a quadratic algebra with commutation relations given by
\begin{align}
 \big[\hat{\Lambda}_{ij}, \hat{\Lambda}_{kl} \big]&  =\imath \hbar \left(\delta_{ik}\big(\hat{\mathds{1}}+2\mu_k \hr_k \big) \hat{\Lambda}_{jl}+\delta_{jl}\big(\hat{\mathds{1}}+2\mu_l \hr_l \big) \hat{\Lambda}_{ik}-\delta_{il}\big(\hat{\mathds{1}}+2\mu_i \hr_i \big) \hat{\Lambda}_{jk}-\delta_{jk}\big(\hat{\mathds{1}}+2\mu_j \hr_j \big) \hat{\Lambda}_{il}\right) \label{A2} \\[2pt]
\big[ \hat{\Lambda}_{ij} ,\hat{\Gamma}_k^{(\kappa)} \big] &=\imath \hbar \left( \delta_{ik}\big(\hat{\mathds{1}}+2\mu_i\hat{R}_i \big)\hat{\Gamma}_j^{(\kappa)} - \delta_{jk} \big(\hat{\mathds{1}}+2\mu_j\hat{R}_j \big)\hat{\Gamma}_i^{(\kappa)}\right)\label{A3}\\[2pt]
\big[\hat{\Gamma}_i^{(\kappa)}, \hat{\Gamma}_j^{(\kappa)} \big]&=4 \imath \hbar \kappa \hat{\Lambda}_{ij} \label{A4} \\[2pt]
 \big[\hat{\Gamma}_i^{(\kappa)}, \hr_{j} \big]&=2 \delta_{ij}\hat{\Gamma}_j^{(\kappa)} \hr_{j} \label{A5} \\[2pt]
 \big[\hat{\Lambda}_{ij}, \hr_{k} \big]&=2 \delta_{ik}\hat{\Lambda}_{ij}\hr_k+2 \delta_{jk}\hat{\Lambda}_{ik}\hr_j \label{A6}\\[2pt]
 \big[ \hr_{i}, \hr_{j} \big]&=0 \, . \label{A7}
\end{align}
The non-zero commutators  in (\ref{A5}) and  (\ref{A6}) can alternatively be expressed in terms of anticommutators (see Remark~\ref{rem1}):
\be
\begin{aligned}
\big\{ \hat{\Gamma}_i^{(\kappa)}, \hr_i  \big\}&= 0 \, , \quad   \big[\hat{\Gamma}_i^{(\kappa)}, \hr_{j} \big] =0\, , \quad i\ne j \\[2pt]
\big\{ \hat{\Lambda}_{ij}, \hr_i  \big\}&=\big\{ \hat{\Lambda}_{ij},\hr_j  \big\}=0\, , \quad \big[ \hat{\Lambda}_{ij} ,\hr_k \big]=0\, ,\quad i\ne j\ne k\, .
\end{aligned}
\label{A8}
 \ee 
 
 \noindent The striking point is that the commutators (\ref{A2})--(\ref{A4}) show that the $N$ operators $\hat{\Gamma}_i^{(\kappa)}$ are proper (curved) momenta on $\bS^N$ and  $\bH^N$. In fact, 
 in the reflection-free  situation, all the commutators (\ref{A2})--(\ref{A7})   simply reduce  to
 \begin{align}
 \big[\hat{\Lambda}_{ij}, \hat{\Lambda}_{kl} \big]&  =\imath \hbar \left(\delta_{ik} \hat{\Lambda}_{jl}+\delta_{jl}  \hat{\Lambda}_{ik}
 -\delta_{il}  \hat{\Lambda}_{jk}-\delta_{jk} \hat{\Lambda}_{il}\right) \nonumber \\[2pt]
  \big[ \hat{\Lambda}_{ij} ,\hat{\Gamma}_k^{(\kappa)} \big] &=\imath \hbar \left( \delta_{ik} \hat{\Gamma}_j^{(\kappa)} - \delta_{jk} \hat{\Gamma}_i^{(\kappa)}\right)\label{A9}\\[2pt]
\big[\hat{\Gamma}_i^{(\kappa)}, \hat{\Gamma}_j^{(\kappa)} \big]&=4 \imath \hbar \kappa \hat{\Lambda}_{ij}   \, ,
\nonumber
\end{align}
which determine the Lie algebra $\mathfrak{so}(N+1)$ for $\kappa>0$, $\mathfrak{so}(N,1)$ for $\kappa<0$, and the (inhomogeneous) Euclidean algebra $\mathfrak{e}(N)\equiv \mathfrak{iso}(N)$ when $
\kappa=0$. Recall that, in this algebraic context, the curvature $\kappa$  plays the role of a graded contraction parameter coming from the $\mathbb Z_2$-grading of (\ref{A9}), provided by the following involutive automorphism (see, e.g.,~\cite{BHSZG} for details) 
\be
\Theta\big(\hat{\Lambda}_{ij},\hat{\Gamma}_i^{(\kappa)}  \big)\to \big(\hat{\Lambda}_{ij},-\hat{\Gamma}_i^{(\kappa)}  \big)\, ,\quad \forall i, j\, .
  \ee
  The change of the sign of $\kappa$ in (\ref{A9}) entails a change of real form of the Lie algebra, while    $\kappa=0$ corresponds to an In\"on\"u--Wigner contraction, which  in this case  leads to the contraction around the origin in the underlying space,  $\bS^N\rightarrow \bE^N\leftarrow \bH^N$. Hence, 
   the set of these three Lie algebras is denoted $\mathfrak{so}_\kappa(N+1)$. Likewise, the automorphism $\Theta$ can be extended to cover the complete quadratic algebra (\ref{A2})--(\ref{A7})
 as
 \be
\Theta\big(\hat{\Lambda}_{ij},\hat{\Gamma}_i^{(\kappa)} ,\hr_i \big)\to \big(\hat{\Lambda}_{ij},-\hat{\Gamma}_i^{(\kappa)}  ,\hr_i \big) \, ,\quad \forall i, j\, ,
 \ee
  and, consistently, the curvature  $\kappa$ only appears in the commutators (\ref{A4}), which  are therefore the only ones that vanish under the flat contraction $\kappa=0$. As far as the realization  (\ref{A1})  is concerned, the curved momenta (\ref{A1}) collapse to the usual (flat) Dunkl momenta $\hat{\Gamma}_i^{(0)}=\hpi_i$.

\noindent Consequently,  by taking into account the  discussion above and  the
    notation introduced in \cite{Ghaz2021} for the subalgebra spanned by the $N(N+1)$ operators  $ \hat{\Lambda}_{ij} $ and $ \hat{R}_i$, we denote by
 \be
 \mathfrak{so}_{\kappa}^{\mu_1, \dots, \mu_N}(N+1):=\mathfrak{so}_{\kappa}\big(N+1; \mu_1 \hat{R}_1, \dots,  \mu_N \hat{R}_N\big) 
 \label{A12}
 \ee
the family of the three quadratic algebras  (\ref{A2})--(\ref{A7}), thus comprising the spherical $(\kappa>0)$, Euclidean $(\kappa=0)$ and hyperbolic cases $(\kappa<0)$, namely
\begin{align}
&\mathfrak{so}_{+}^{\mu_1, \dots, \mu_N}(N+1):=\mathfrak{so}\big(N+1; \mu_1 \hat{R}_1, \dots,  \mu_N \hat{R}_N \big) \nonumber  \\
&\mathfrak{so}_{0}^{\mu_1, \dots, \mu_N}(N):=\mathfrak{e}\big(N; \mu_1 \hat{R}_1, \dots,  \mu_N \hat{R}_N \big)  \label{A13}\\
 &\mathfrak{so}_{-}^{\mu_1, \dots, \mu_N}(N,1):=\mathfrak{so} \big(N,1; \mu_1 \hat{R}_1, \dots,  \mu_N \hat{R}_N \big)  \, .
 \nonumber
\end{align}
Then, the quadratic algebra (\ref{A12}) is an extension by reflections of the  $N(N+1)/2$-dimensional Lie algebra $\mathfrak{so}_{\kappa}(N+1)$ (\ref{A9}). Finally, note that the three particular cases (\ref{A13}) contain the same  $N(N+1)/2$-dimensional subalgebra  $\mathfrak{so}\big(N, \mu_1 \hr_1, \dots, \mu_N  \hr_N\big)$  in \cite{Ghaz2021}, 
 fulfilling the commutation rules (\ref{A2}),  (\ref{A6}) and  (\ref{A7}).


\end{document}